\documentclass[useAMS,usenatbib,structabstract]{mn2e}

\usepackage{graphicx,latexsym}
\usepackage{color}
\usepackage{natbib}
\usepackage{rotating,ams}
\usepackage{appendix}
\usepackage[ps2pdf,linktocpage]{hyperref}
\hypersetup{%
  colorlinks=false,
  bookmarksopen=true,
  bookmarksnumbered=true,
  pdfstartpage={1},  
  pdftitle = {How well do {\sc starlab} and {\sc nbody6} compare?
  II: Hardware and accuracy},
  pdfauthor = {P. Anders, H. Baumgardt, E. Gaburov, S. Portegies Zwart} ,
  pdfkeywords = {N-body simulations},
  pdfproducer = {LaTeX with hyperref}
}

\newcommand{\be}{\begin{enumerate}} 
\newcommand{\ee}{\end{enumerate}} 
\newcommand{\bi}{\begin{itemize}} 
\newcommand{\ei}{\end{itemize}} 

\def\spose#1{\hbox to 0pt{#1\hss}}
\def\lta{\mathrel{\spose{\lower 3pt\hbox{$\mathchar"218$}}
     \raise 2.0pt\hbox{$\mathchar"13C$}}}
\def\gta{\mathrel{\spose{\lower 3pt\hbox{$\mathchar"218$}}
     \raise 2.0pt\hbox{$\mathchar"13E$}}}

\newcommand{\aap}{A\&A}

\newcommand{\apjl}{ApJ}

\newcommand{\apjs}{ApJS}

\title[Comparison {\sc nbody} vs. {\sc starlab} II.]{How well do {\sc
starlab} and {\sc nbody} compare? II: Hardware and accuracy}

\author[P. Anders et al.]{P. Anders $^{1}$ \thanks{E-mail: anders@pku.edu.cn}, 
H. Baumgardt $^{2,3}$, E. Gaburov $^{4,5}$, S. Portegies Zwart$^{6}$\\
$^{1}$ Kavli Institute for Astronomy and Astrophysics, Peking University, Yi He Yuan Lu 5, Hai Dian District, Beijing 100871, China\\
$^{2}$ Argelander Institut f\"ur Astronomie, Universit\"at Bonn, Auf dem H\"ugel 71, 53121 Bonn, Germany\\
$^{3}$ School of Mathematics and Physics, The University of Queensland, Brisbane, QLD 4072, Australia\\
$^{4}$ Center for Interdisciplinary Exploration and Research in Astrophysics (CIERA) \& Dept. of Physics and Astronomy, \\
Northwestern University, 2145 Sheridan Rd, Evanston, IL 60208, USA\\
$^{5}$ Hubble fellow\\
$^{6}$ Sterrewacht Leiden, Universiteit Leiden, Niels Bohrweg 2, 2333 CA Leiden, The Netherlands\\
}
\begin{document}


\date{Accepted ---. Received ---; in original form ---.}
\pubyear{2011}
\maketitle

\begin{abstract}

Most recent progress in understanding the dynamical evolution of star
clusters relies on direct $N$-body simulations. Owing to the
computational demands, and the desire to model more complex and more massive
star clusters, hardware calculational accelerators, such as GRAPE special-purpose
hardware or, more recently, GPUs (i.e. graphics cards), are generally
utilised. In addition, simulations can be accelerated by adjusting
parameters determining the calculation accuracy (i.e. changing the
internal simulation time step used for each star).

We extend our previous thorough comparison (\citealt{2009MNRAS.395.2304A}) of
basic quantities as derived from simulations performed either with {\sc
starlab/kira} or {\sc nbody6}. Here we focus on  differences arising from using
different hardware accelerations (including the increasingly popular graphic
card accelerations/GPUs) and different calculation accuracy settings.

We use the large number of star cluster models (for a fixed stellar mass
function, without stellar/binary evolution, primordial binaries,
external tidal fields etc) already used in the previous paper, evolve
them with {\sc starlab/kira} (and {\sc nbody6}, where required), analyse
them in a consistent way and  compare the averaged results
quantitatively. For this quantitative comparison, we apply the bootstrap
algorithm for functional dependencies developed in our previous study.

In general we find very high comparability of the simulation results, 
independent of the used computer hardware (including the hardware accelerators)
and the used $N$-body code. For the tested accuracy settings we find that for reduced
accuracy (i.e. time step at least a factor 2.5 larger than the standard setting)  most
simulation results deviate significantly from the results using standard
settings. The remaining deviations are comprehensible and explicable.

\end{abstract}


\section{Introduction}

In recent years, the modelling of star populations strongly advanced due to stellar dynamical
studies. The fields covered comprehend a wide range of research questions, such as the modelling of
individual star clusters (e.g.
\citealt{2003ApJ...589L..25B,2005MNRAS.363..293H,2009MNRAS.397L..46H,2010MNRAS.409..628H,2011MNRAS.411.1989Z}), 
star cluster systems (e.g. \citealt{2003ApJ...593..760V}), populations of ``exotic'' stellar objects
(\citealt{2004Natur.428..724P} - massive black holes, \citealt{2008ApJ...680L.113U} - blue
stragglers,  \citealt{2010A&A...516A..73D} - second populations of globular cluster stars,
\citealt{2005MNRAS.363..223G,2008MNRAS.385..929G,2011arXiv1111.3644F} - run-away stars) etc.

The major codes used in this field are the family of {\sc nbodyx} codes
(\citealt{1999PASP..111.1333A}) and the {\sc starlab} environment with its $N$-body
integrator {\sc kira} (\citealt{2001MNRAS.321..199P}). Both codes are continuously expanded and improved, 
including the ability for the use of hardware dedicated to improve the simulation capabilities
(in terms of simulation duration and available population size). This includes 
advances in software development (such as parallelisation, see e.g. 
\citealt{2007NewA...12..357H,2008LNP...760..377S,2008NewA...13..285P}) and the use of dedicated hardware, e.g. 
GRAPEs (\citealt{1993PASJ...45..269E,2003PASJ...55.1163M}), more recently the use of 
GPUs (\citealt{2007NewA...12..641P,2009NewA...14..630G,gpunbody,2011NewA...16..445M}), 
supercomputers/computer clusters (\citealt{2007NewA...12..357H}) and large area networks, such as GRID 
(\citealt{2011CS&D....4a5001G}).  

Additional $N$-body codes are emerging, such as MYRIAD (\citealt{2010A&A...522A..70K}, GRAPE-enabled),
NBSymple (\citealt{2011NewA...16..284C}, GPU-enabled) and AMUSE\footnote{Available at
http://www.amusecode.org/} (\citealt{2009NewA...14..369P}, a compilation of various modules for
gravitational dynamics, stellar evolution, radiative transfer and hydrodynamics, which can become
independently combined). GPUs are also applied in an additional wide range of astrophysical simulation
environments (\citealt{2009NewA...14..406F,2010ApJS..186..457S,2010NewA...15...16T}). Their general
availability and useability for computationally intensive astrophysical modelling will continue the rise
of GPU usage.

Despite their extensive use and computational specialities (see Sects. \ref{sec:hardwaresoftware} and \ref{sec:acc}), a detailed study concerning the reliability and comparability of simulations performed is missing. Important factors,
which are studied in the present work, include the use of various computer hardware (including diverse
special-purpose hardware to accelerate the simulations) and simulation accuracy settings. The named factors determine
the used calculation accuracy and therefore the accuracy to trace accurately fast important phases of star cluster
evolution. Potentially, this can lead to incommensurable simulation results, although starting configuration and physical
treatment are equivalent. 

In Sect. \ref{sec:hardwaresoftware} describes the used hardware and software, including their
limitations in calculation accuracy. Sect. \ref{sec:methods} summarizes the used simulation and
analysis methods. In Sect. \ref{sec:results} the simulation results are explained and analysed,
with a summary of all results and drawn conclusions in Sect. \ref{sec:summary}.

\section{Utilised hardware and software}
\label{sec:hardwaresoftware}

We used 3 PCs for these simulations, whose specifications are summarised in
Table \ref{tab:computers}. The use of these 3 PCs was necessary to allow testing
the sensitivity of the results on the PC architecture and the availability of
varying accelerator hardware. The labels ``PC1'', ``PC2'' and ``PC3'' will be
used to indicate the use of solely the CPU, without the available accelerator
hardware. ``PC1/GPU'', ``PC2/GRAPE'' and ``PC3/GPU'' indicate the usage of the
available accelerator hardware.

\begin{table*}

\caption{Specifications of PCs used} 

\begin{center}
\begin{tabular}{l c c c}
\hline
specification & PC1 & PC2 & PC3\\
\hline
Processor  & Quad Core Intel Q9300 & Dual Core Intel Xeon &  Dual Core AMD Opteron  \\
Operating system  & Debian 4.0 & Red Hat 2.6.9-67 & Debian 2.6.24.7   \\
C compiler  & gcc 4.2 & gcc 3.4.6 &  gcc 4.3.2  \\
Special hardware  & GeForce 9800 GX2 & GRAPE6-BLX64 &  GeForce 9800 GX2  \\
Special hardware libraries  & Sapporo v1.5 & g6bx-0515 & GPU module by K. Nitadori  \\
$N$-body software & {\sc starlab} 4.4.4 & {\sc starlab} 4.4.4 & {\sc nbody6} \\
\hline
\end{tabular}
\label{tab:computers}
\end{center}
\end{table*}

For the majority of tests we use the {\sc starlab} software package. Where we
supplement these with simulations made with  {\sc nbody} we use {\sc nbody6},
contrary to our previous study (\citealt{2009MNRAS.395.2304A}, hereafter ``Paper
I'') where we used {\sc nbody4}. {\sc nbody6} presents a significant
enhancement of required computational time due to the AC neighbour scheme
(\citealt{1973JCoPh..12..389A} for the AC neighbour scheme and
\citealt{2001NewA....6..277A} for the {\sc nbody} implementation).

\subsection{Possible reasons for simulation differences}

Numerical modeling of $N$-body systems, such as star clusters, usually involves
long-term simulations. It is therefore necessary to understand the influence of
different errors on the outcome. In a very basic case of a direct integration,
there are two contributions to the overall error: one due to the time integration
and one due to the accuracy of force calculations. In the idealistic case these
two should be balanced, but in the majority of codes (e.g. {\sc nbody4}, {\sc
starlab}, phiGrape \citealt{2008MNRAS.389....2H}, Myriad \citealt{2010A&A...522A..70K}) 
a simple direct summation method is employed to
compute forces, whereas {\sc nbody6} uses the Ahmad-Cohen neighbour scheme.
While this, in principle, should result in exact accelerations, in practice, due
to limited precision, several factors affect the result. For example, the order in
which partial forces are added affects the accuracy: if the partial forces are
added from the weakest to strongest the total force on a particle is more accurate
compared to the one if the summation is done in reverse or in arbitrary order.
This is exacerbated if the precision of the force accumulation is below a certain
threshold which depends on the system being modelled. On CPU, a generic double
precision force loop usually adds up partial forces in IEEE double precision
arithmetics, which, is safe to say, results in the most accurate force. On special
purpose accelerations (such as GRAPE or GPU), for performance considerations, the
summations is done in lower  precision (e.g.
\citealt{2003PASJ...55.1163M,NitadoriThesis,2009NewA...14..630G}). For example,
\citet{NitadoriThesis} and \citet{2009NewA...14..630G} use two single precision
numbers to emulate double precision, and the second authors conduct force
accumulation in single precision. The emulation of double precision is inherently
less precise (2x23 bits mantissa) compared to IEEE double precision (53 bits
mantissa), and its long-term errors are poorly understood. This may potentially
inflict damage to long-term integrations.

In direct $N$-body simulations the error is usually due to time integration (among
other errors coming from binary and multiple treatment, stellar evolution, close
encounter treatment) rather than force summation. Generally in $N$-body codes, the
time-step is defined by the Aarseth criterion (\citealt{2003gnbs.book.....A})
which is usually scaled by an ``accuracy parameter'' $\eta$. The time-step size
monotonically depends on $\eta$ and changes in this parameter have proportional
changes in the value of the time-step. In a generally employed $4^{th}$ order
Hermite scheme, the error decreases as the $4^{th}$ order of the time-step, and
for a realistic time-step criterium which provides a good balance between duration
of simulations and accuracy, this error remains dominant. Furthermore, time
integration error has similar properties among different codes due to its simpler
nature. 

If $\eta$ is increased above a certain value, the time integration errors become
intolerable, for example close encounters and dynamical binary integration become
inaccurately simulated, but consistent among different codes. If $\eta$ is
decreases beyond a certain value, the dominant source of error becomes force
computation, which varies among different implementations to a large degree. While
this produces more accurate results, the outcome from different codes may show
large scatter in sensitive quantities, such as energy conservation error, due to
possibly significantly different force summation (or other) errors, rather than
similar time integration errors, influence long-term simulations. In this paper we
explore a range of $\eta$ parameters to identify their influence on the simulation
results.

\section{Methods}
\label{sec:methods}

In this section we will summarize the methods developed in
Paper I and applied in the present paper. For
additional details see Paper I.

\subsection{Standard setup}

We used start models from the predecessor study. They were created using {\sc
starlab} (and, where necessary, transformed into the {\sc nbody} file format)
and have the following characteristics

\bi
\item 1024 (1k) particles
\item \citet{1955ApJ...121..161S} stellar mass function, with the
highest mass = 10 $\times$ lowest mass
\item no primordial binaries
\item \citet{1911MNRAS..71..460P} sphere initial density profile
\ei 

There are 50 start models, with identical parameters but varying statistical
realisations.\footnote{The start models are available at
http://data.galev.org/nbody/NBODY\_STARLAB\_Comparison/ as ``MF10 models'' and 
at http://www.initialconditions.org/14.}

All simulations were performed with the following settings:

\bi
\item no stellar/binary star evolution
\item no external tidal field
\item duration of simulation: 1000 $N$-body time units (i.e., until well
after core collapse, which occurs around 60 $N$-body time units, see
Paper I).
\ei

Unless otherwise specified, the simulations were performed using

\bi
\item hardware acceleration: GeForce 9800GX2 GPU
\item calculation accuracy parameter $\eta$=$\eta_{\rm STARLAB}$=0.1 for {\sc
starlab}, equivalent to $\eta_{\rm NBODY}$=0.01 (due to a different definition
of the accuracy parameter in {\sc starlab} and in {\sc nbody6}, see Sect.
\ref{sec:acc}). All quoted accuracy parameters are given as 
$\eta$=$\eta_{\rm STARLAB}$.
\ei

\subsection{Analysis of $N$-body simulation snapshots}

As we want to trace differential effects between our settings (i.e., {\sc
nbody6} vs. {\sc starlab}, solely CPU vs. CPU+GRAPE vs. CPU+GPU, different
accuracy parameters) we need to ensure that output from different simulations is
treated identically. The main difficulty arises when comparing {\sc nbody6} and
{\sc starlab} output, as the output content and format is very different. E.g.,
{\sc nbody6} output does not contain detailed information about binaries, while
{\sc starlab} output contains information on binary/triple/higher order
hierarchies.

We therefore reduce the format of each simulation output to the most
basic one possible, only containing for each star its mass and 
vectors for its position and velocity. Binaries and higher order
hierarchies are split up into their components. Other possibly present
information is removed.

We then split the series of output into individual snapshots,
pass them through {\sc kira}, the $N$-body integrator of the
{\sc starlab} package, which evolves them for the 1/32$^{th}$ part of
an $N$-body time unit. This procedure rebuilds the binary/higher order
hierarchy structure, and stores it accessibly in the resulting snapshot.
All snapshots with rebuilt hierarchy structure are then analysed using
tools\footnote{Specifically the {\sc hsys\_stats} program of {\sc
starlab}.} available within the {\sc starlab} package. In Paper I we 
investigated the reliability of this
procedure and did not find significant biases (except for the spurious
re-building of multiple systems in the outer cluster regions, which are
not subject of the present study). In addition, we calculate the change of 
total energy per N-body time unit as a measure of the energy conservation.

Results of this procedure are time series of cluster parameters (like
core radius, virial ratio, etc) and binary parameters (here only
investigated at the end of the simulation; as shown in Paper I testing at
different times gives comparable results but poorer statistics). For the
further analysis, results from the 50 individual runs per simulation
setting are combined: the lists of binary parameters are merged, while
for the overall cluster parameters a median time series and the 16/84
per cent quantiles are calculated.

\subsection{Analyses of the simulation outcome}

In this section we want to summarize the methods used for the analysis
of the $N$-body data. Many of the analysis
results are given as probabilities. We will adopt the following (commonly
used) nomenclature when discussing the significance level of 
statistical test results

\begin{itemize} 
\item {\sl highly significant}: p-value $<$ 1\% 
\item {\sl significant}: p-value $<$ 5\% 
\item {\sl weakly significant}: p-value $<$ 10\%.
\end{itemize}

``Multiple testing'': if we perform 100 independent tests, a fraction of the
order of $10\%$ of p-values below $0.1$ will arise by chance even if none of the
test null hypotheses is wrong. As we analyse various star cluster parameters of
a large number of simulations, we need to take ``multiple testing'' into
account.

\subsubsection{Binary and escaper parameters}

We will analyse the parameters of dynamically created binaries
(semi-major axis, eccentricity, mass ratio, binding energy) and the
energy distributions of stars escaping from the cluster.

Differences between two such distributions for two different simulation
setups are evaluated using a Kuiper test (\citealt{Kuipertest}), an
advanced KS test (for the KS test and Kuiper test see, e.g., Numerical Recipes
\citealt{1992nrfa.book.....P}). A Kuiper test returns the probability
that two distributions are drawn from the same parent distribution (or,
more accurately, the probability that one is wrongly rejecting the null
hypothesis ``The two distributions are drawn from the same parent
distribution.'', which in most cases is equivalent).

\subsubsection{Time series}
\label{sec:methods.timeseries}

In Paper I we developed a bootstrap algorithm for
functional dependencies and applied it to time series similar to the ones
studied in the present work. For example, we want to quantitatively
compare the time evolution of the core radius (and of other parameters)
for two different simulations setups.

First, we introduce a measure of the difference between two time series'
(or, more general, between two functional dependencies). Assume we have 
two functional dependencies of one parameter from the
independent variable $x$: $y_1(x)$ and $y_2(x)$. For each $x$ these
dependencies have uncertainties $\sigma_1(x)$ and $\sigma_2(x)$. The 
relative difference between the functional dependencies at a given
$x$ is then:

\begin{equation}
\delta_{12}(x) = \frac{y_1(x) -
y_2(x)}{\sqrt{\sigma_1(x)^2+\sigma_2(x)^2}}
\label{eq:single}
\end{equation}

We then define the ``difference between functions 1 and 2'' as 

\begin{equation}
\Delta_{12} = \frac{1}{N} \cdot \left | \sum_x \delta_{12}(x) \right |
\label{eq:sum1}
\end{equation}

where $N$ is the number of datapoints used for the statistic. We
consider only the absolute value, as we want to have a measure of the
{\sl size} of the difference, but not necessarily its {\sl direction}.
In addition, this ensures $\Delta_{12} \equiv \Delta_{21}$.

Equivalently we define the ``{\sl absolute} difference between
functions 1 and 2'' as 

\begin{equation}
\Gamma_{12} = \frac{1}{N} \cdot  \sum_x \left | \delta_{12}(x) \right |
\label{eq:sum2}
\end{equation}

While $\Delta_{12}$ is more sensitive to systematic offsets,
$\Gamma_{12}$ traces also statistical fluctuations.

We then utilise this measure of difference between two time series' to
design a bootstrap-like algorithm to quantify the significance of this
difference.

We calculate 300 test clusters with {\sc starlab} using the standard
settings (i.e., using PC1/GPU and with $\eta$=0.1), and the same
analysis routines as for the other clusters. 

From these test clusters we randomly select sets of 50 clusters each
(i.e. the number of clusters in the main simulations) {\bf with
replacement}, and calculate for each parameter the median $y^T(x)$ and
quantiles $\sigma^T(x)$. 

We build 2000 such sets. Out of those we randomly select two sets (again
with replacement) and derive the individual values of $\Delta_{12}^T$
and $\Gamma_{12}^T$. We repeat this procedure 10000 times to estimate
the $\Delta_{12}^T$ and $\Gamma_{12}^T$ test distributions for each
parameter. As all test clusters are calculated with the same settings,
the $\Delta_{12}^T$ and $\Gamma_{12}^T$ test distributions represent the
null hypothesis ``functions 1 and 2 are drawn from the same parent
distribution''. By comparing these test distributions with the values
derived from the main simulations $\Delta_{12}^S$ and $\Gamma_{12}^S$ we
can quantify the fraction of data in the test distribution with
$\Delta_{12}^T$ or $\Gamma_{12}^T$ more deviating than the values
derived from the main simulations $\Delta_{12}^S$ or $\Gamma_{12}^S$.
This value serves as measure of how similar the two main simulations
are.

In order to avoid applying the test statistic to highly correlated data,
which appears for the earliest timesteps (as all runs share the same type of 
start models, only different stochastic realisations) and which is beyond the 
area of application of the test
statistic, we start the summation in Eq. \ref{eq:sum1} and \ref{eq:sum2}
at 20 $N$-body time units. This is sufficiently well before core collapse,
which appears at about 60 $N$-body time units (see Paper I), to capture 
the systems' behaviour during this important epoch. As we have shown in 
Paper I, the influence of varying this start time slightly is small and
does not change the conclusions drawn.

\section{Results}
\label{sec:results}

In this section we will compare simulations made for a number of
specifications. 

The quantitative test results are tabulated in Appendix \ref{app:tables} and 
summarized in Sect. \ref{sec:summary}. Visual presentations of time series for 
selected results are provided in Appendix \ref{app:graphics}.
.


\subsection{Different hardware, using {\sc starlab}}

We have performed simulations on PC1 and PC2. For each PC we realised one set of
simulations with the installed accelerator hardware and one set without hardware
acceleration.

\begin{figure*}
\begin{center}
  \begin{tabular}{cc}   
   \includegraphics[angle=270,width=0.37\linewidth]{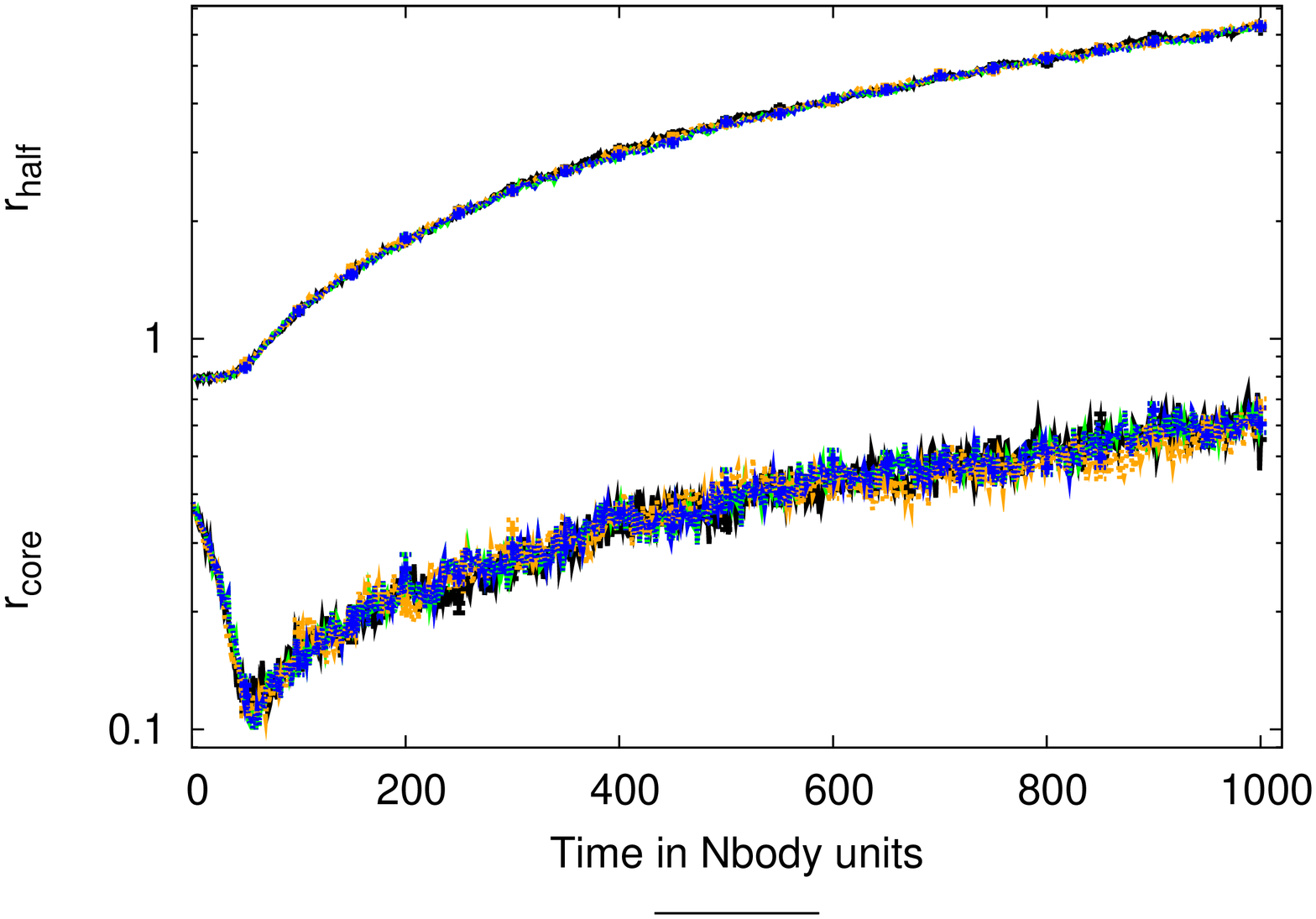} & 
   \includegraphics[angle=270,width=0.37\linewidth]{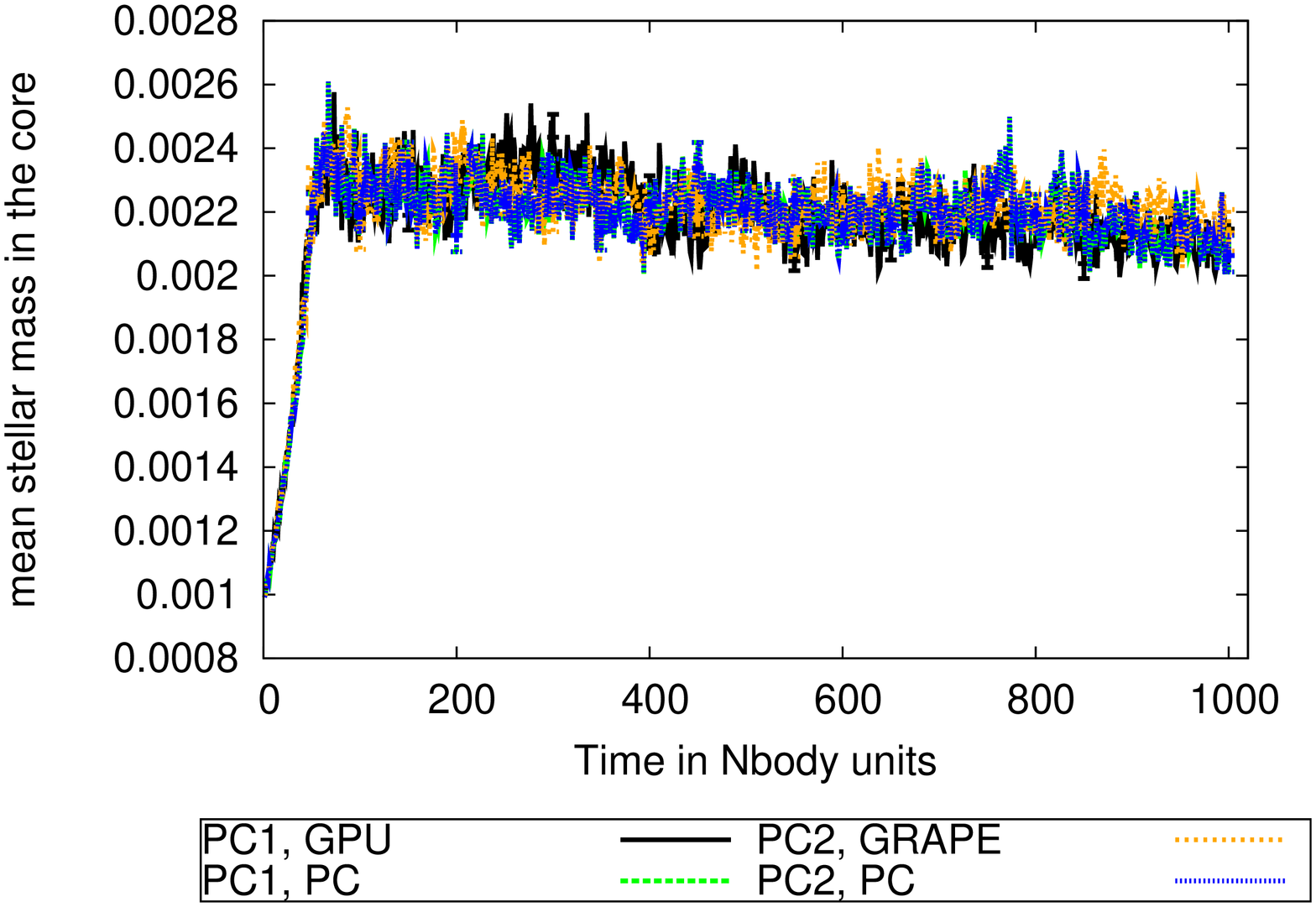} \\
   \includegraphics[angle=270,width=0.37\linewidth]{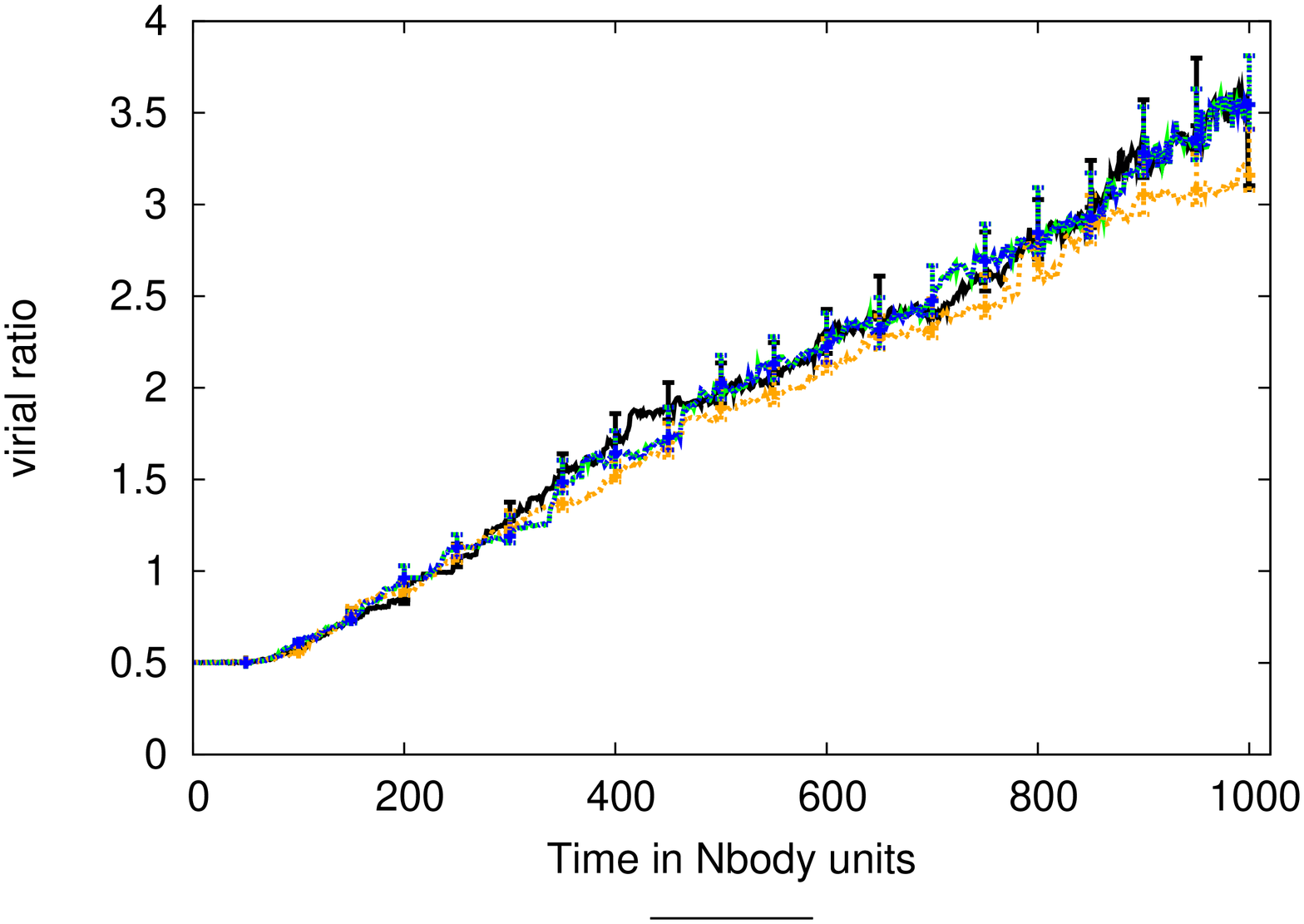} & 
   \includegraphics[angle=270,width=0.37\linewidth]{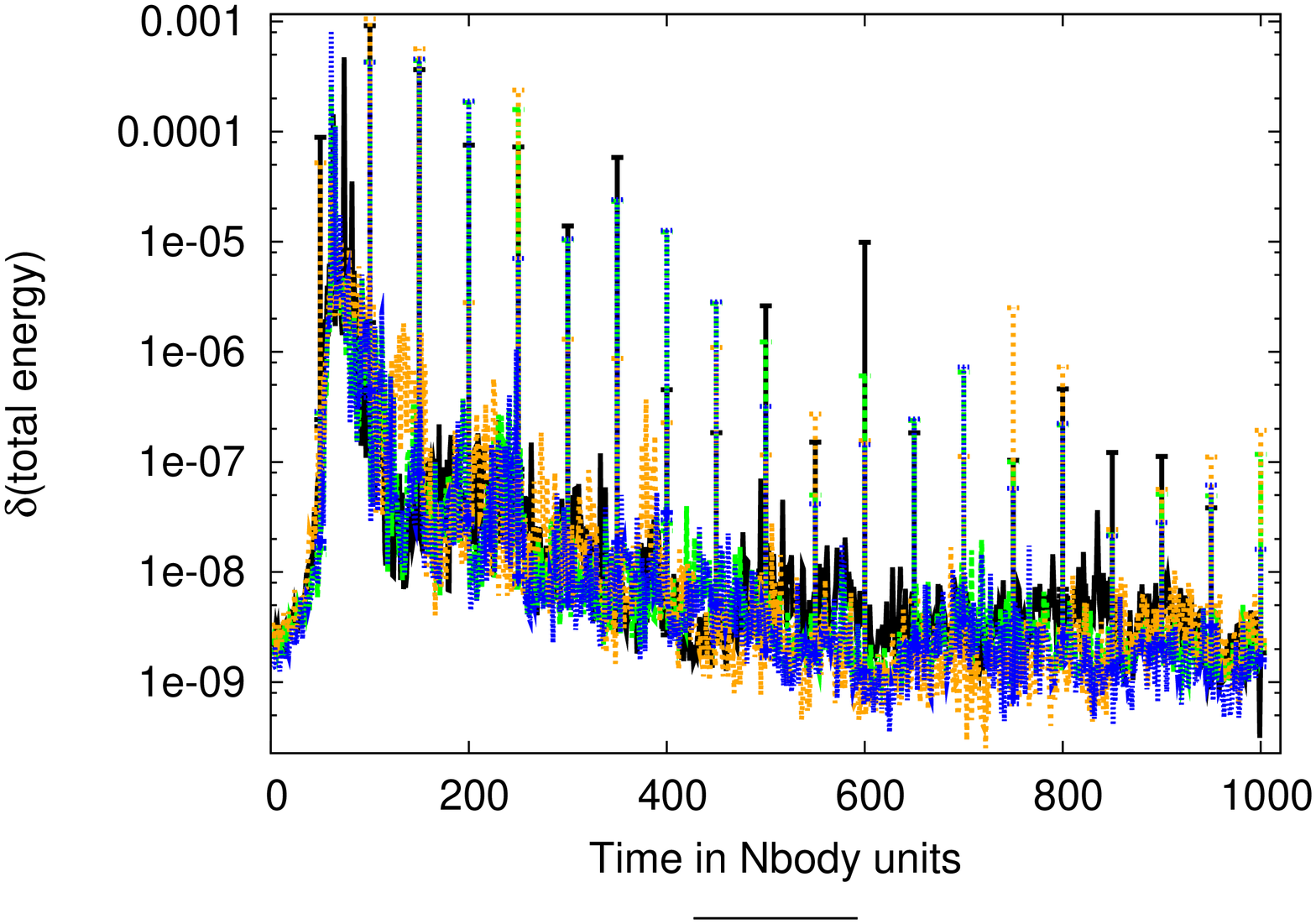} \\
  \end{tabular}
\end{center}

   \caption{Comparison of simulations using {\sc starlab} and various
   hardware.  The lines show the median values, the error bars give the
   uncertainty ranges from the 50 individual runs. Shown are the time
   evolutions of the core radius (top left, bottom lines), half-mass
   radius (top left, upper lines), the objects' mean mass in the core = 
   measure for mass segregation (top right), virial ratio (bottom
   left) and change in total energy during 1 $N$-body time unit (bottom right).}

\label{fig:hw_struct_par}

\begin{center}
   \begin{tabular}{cc}
     \includegraphics[angle=270,width=0.37\linewidth]{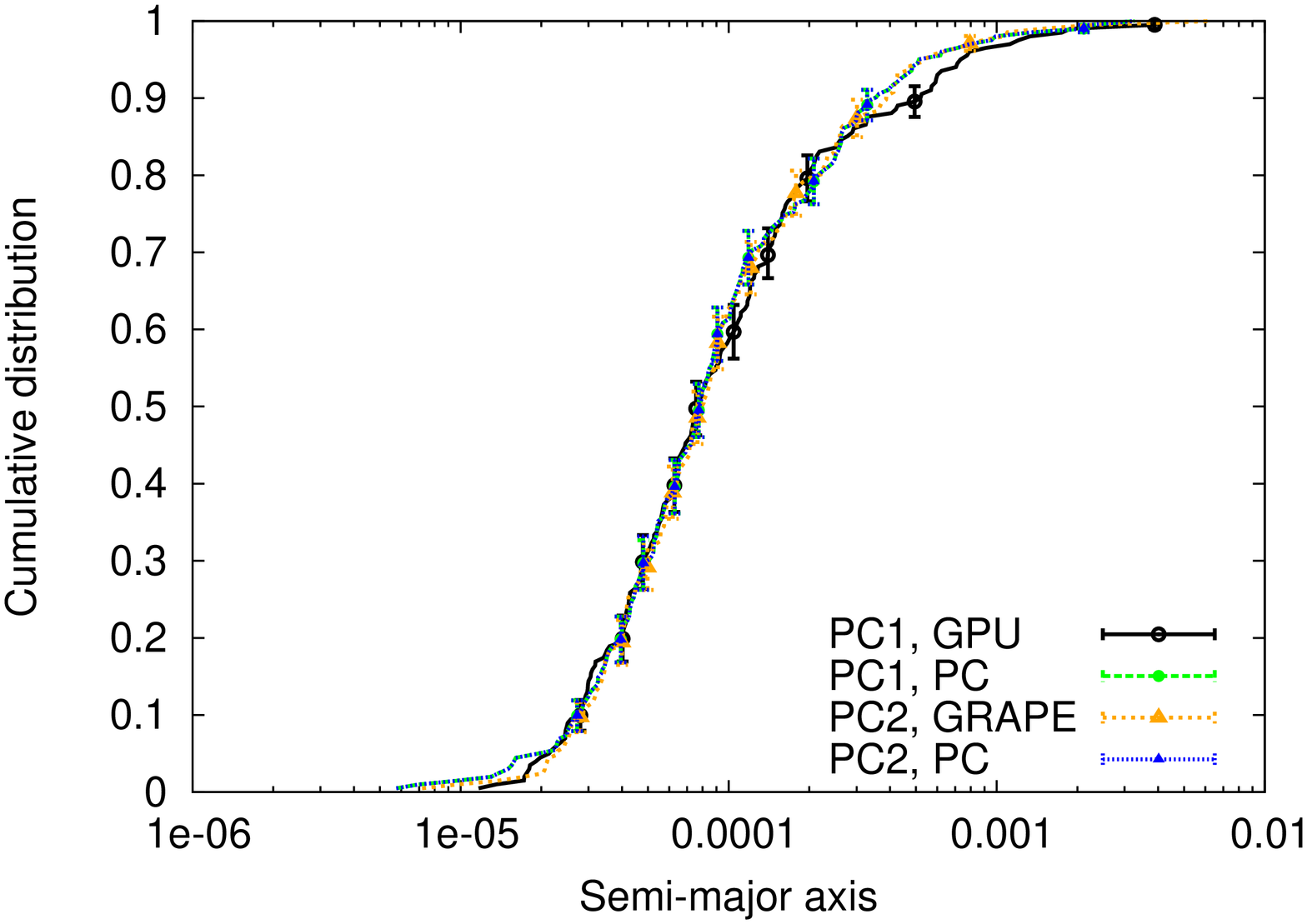} &
     \includegraphics[angle=270,width=0.37\linewidth]{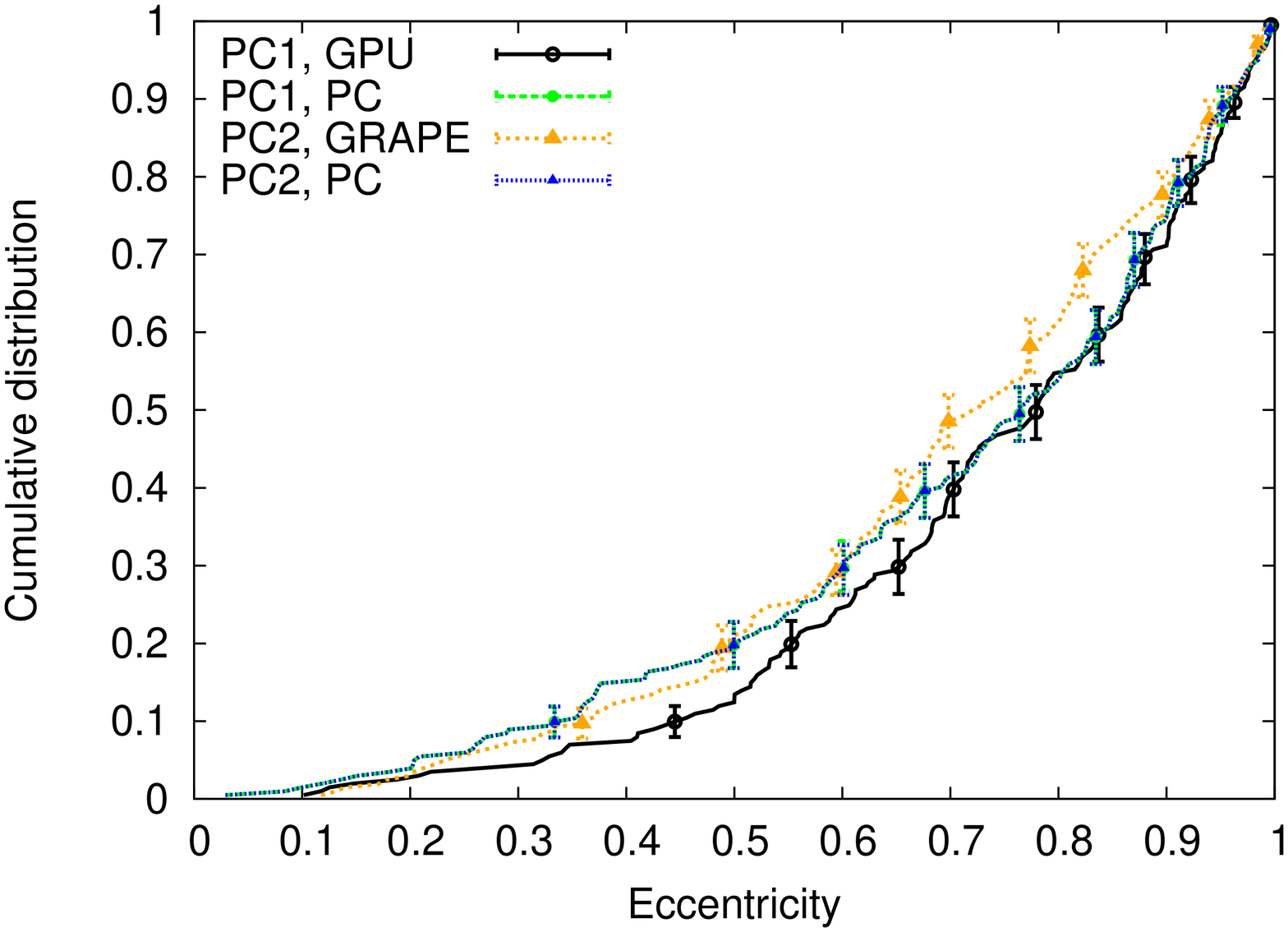} \\
     \includegraphics[angle=270,width=0.37\linewidth]{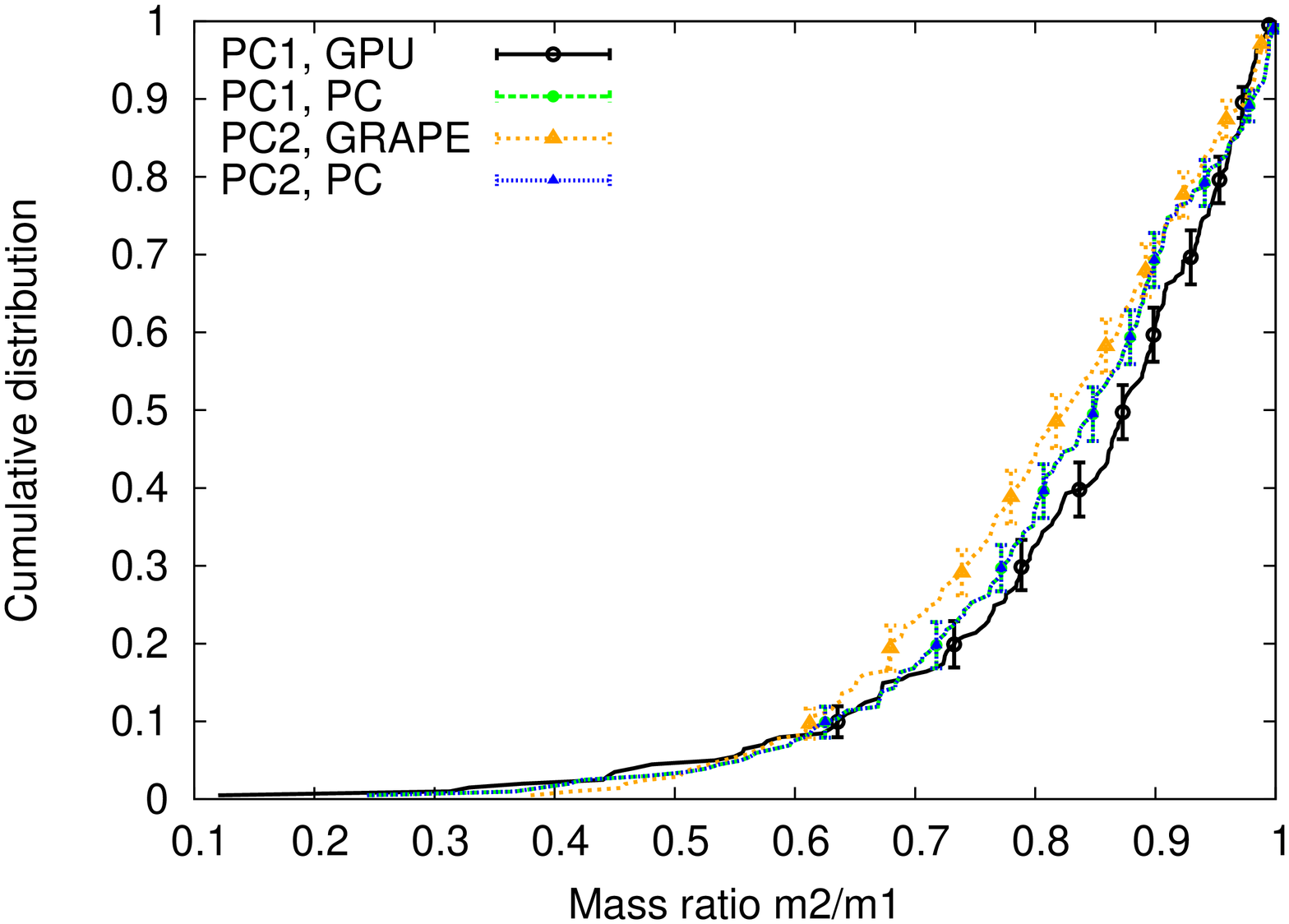} &
     \includegraphics[angle=270,width=0.37\linewidth]{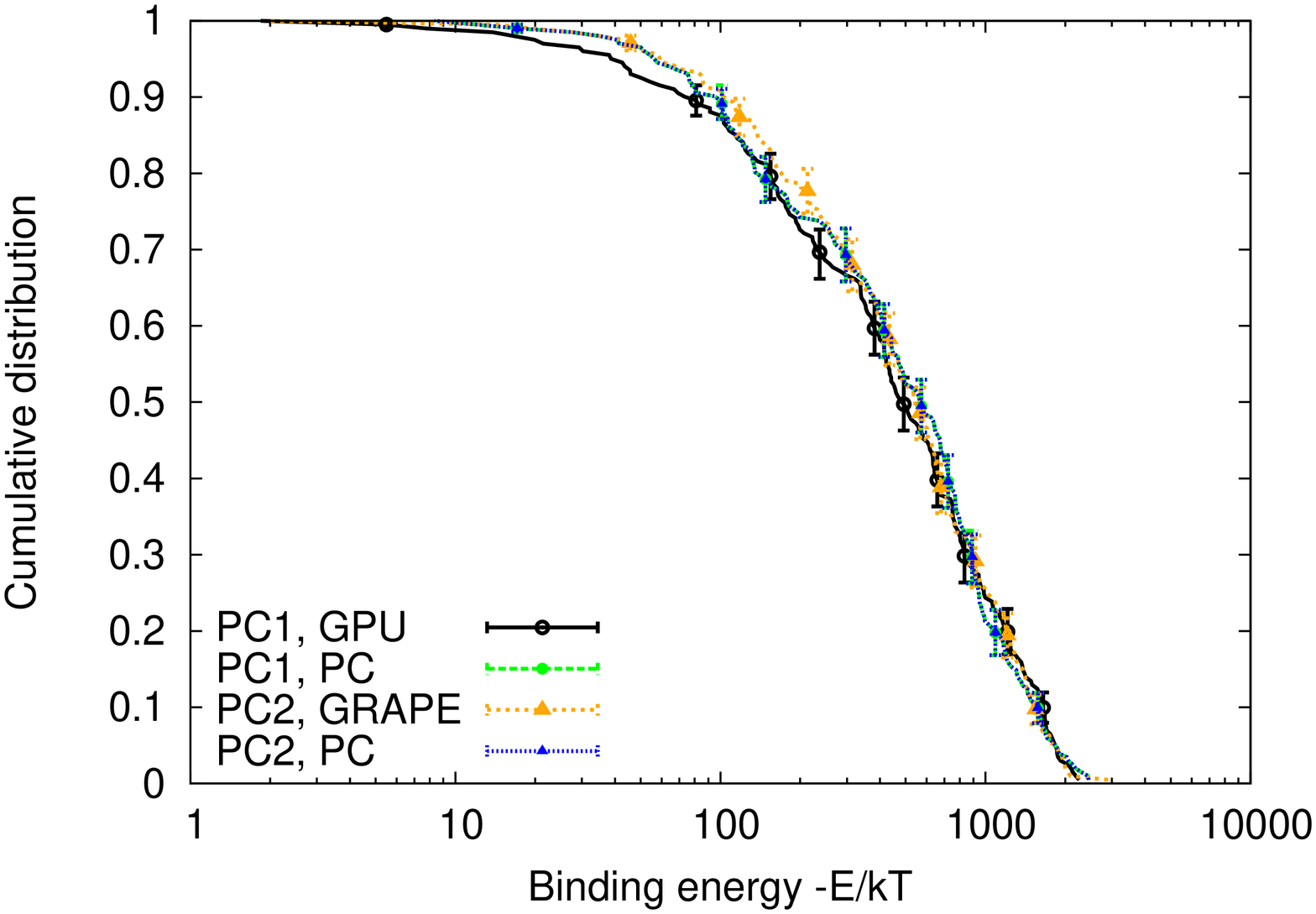} \\
   \end{tabular}

   \caption{Comparison of binary parameters from simulations using  {\sc
   starlab} and various hardware after 1000 $N$-body time units (i.e.,
   well after core collapse). Shown are the cumulative distributions of
   the semi-major axis  (top left), the eccentricity (top right), mass
   ratio of secondary to primary  (bottom left) and the binding energy
   (bottom right). The lines show the data, the error
   bars give the uncertainty ranges from bootstrapping (at every 
   20$^{th}$ data point only, for clarity).}

   \label{fig:hw_bin_par}
\end{center}
\end{figure*}

The results are visualised in Figs. \ref{fig:hw_struct_par} and
\ref{fig:hw_bin_par}. The agreement appears to be very good, with
possible exceptions being the evolution of the kinetic energy (though
with large error bars, only the virial ratio is shown for consistency 
with the other sections) and the cumulative eccentricity and mass ratio
distributions of the dynamically created binaries. The quantitative
measures (our bootstrap algorithm for integrated cluster properties, and
the Kuiper test for the properties of binaries and escaping stars),
shown in Tables \ref{tab:res_boot_struct} - \ref{tab:kstest_esc}, 
generally confirm the good agreement. The following issues are found

\begin{itemize}

\item significant differences: kinetic energy of stars escaping from the
cluster for \\ \#1.1 PC1 $\Leftrightarrow$ PC1/GPU \\
and differences (visible in Fig. \ref{fig:hw_bin_par}) in the mass 
ratio distributions for \\
\#1.1 PC1 $\Leftrightarrow$ PC1/GPU \\
\#1.2 PC1/GPU $\Leftrightarrow$ PC2/GRAPE

\item weakly significant: the $\Delta_{12}$ test of King W$_0$ and the 
potential energy for \\ \#1.3 PC2 $\Leftrightarrow$ PC2/GRAPE

\item statistically insignificant: differences in the total
kinetic energy of all stars and the binary eccentricity distributions (visible in Fig. \ref{fig:hw_bin_par})

\end{itemize}

Deviations in the mass ratio distributions can originate from the stochastical
process of two-body interaction processes. The number and level of the remaining
deviations  is in agreement with ``multiple testing'', i.e. the occurence of
statistical disagreement based on large numbers of tests. Therefore, $N$-body
simulations based on equivalent start conditions and $N$-body code (here {\sc
starlab}) produce highly comparable results, independent of the computer hardware
(including the accelerator hardware). 

\subsection{Accuracy settings}
\label{sec:acc}

Using PC1/GPU, we test the impact of the calculation accuracy. 

The ``calculation accuracy'' is determined by a
dimensionless constant $\eta$ in the formula determining the integration
time step (see \citealt{2003gnbs.book.....A}, Sect. 2.3)

\begin{eqnarray} 
\Delta t_i & = & \sqrt{\eta_{nbody}   \frac{(|\mathbf{F}||\mathbf{F^{(2)}}|+|\mathbf{F}|^2)} { |\mathbf{F^{(1)}}||\mathbf{F^{(3)}}| + |\mathbf{F^{(2)}}|^2 }} \nonumber \\
           & = & \eta_{starlab} \sqrt{\frac{(|\mathbf{F}||\mathbf{F^{(2)}}|+|\mathbf{F}|^2)} { |\mathbf{F^{(1)}}||\mathbf{F^{(3)}}| + |\mathbf{F^{(2)}}|^2 }}
\end{eqnarray}

where $\mathbf{F}$ is the force affecting particle $i$ and 
$\mathbf{F}^{(j)}$ its $j$-th time derivative.

Due to the slightly different definitions used by {\sc starlab} and {\sc nbody}
is $\eta_{\rm starlab}$=$\sqrt{\eta_{\rm nbody}}$. In the remainder of the paper
we will express all accuracy settings as $\eta$=$\eta_{\rm starlab}$.

\begin{figure*}
\begin{center}
  \begin{tabular}{cc}   
   \includegraphics[angle=270,width=0.37\linewidth]{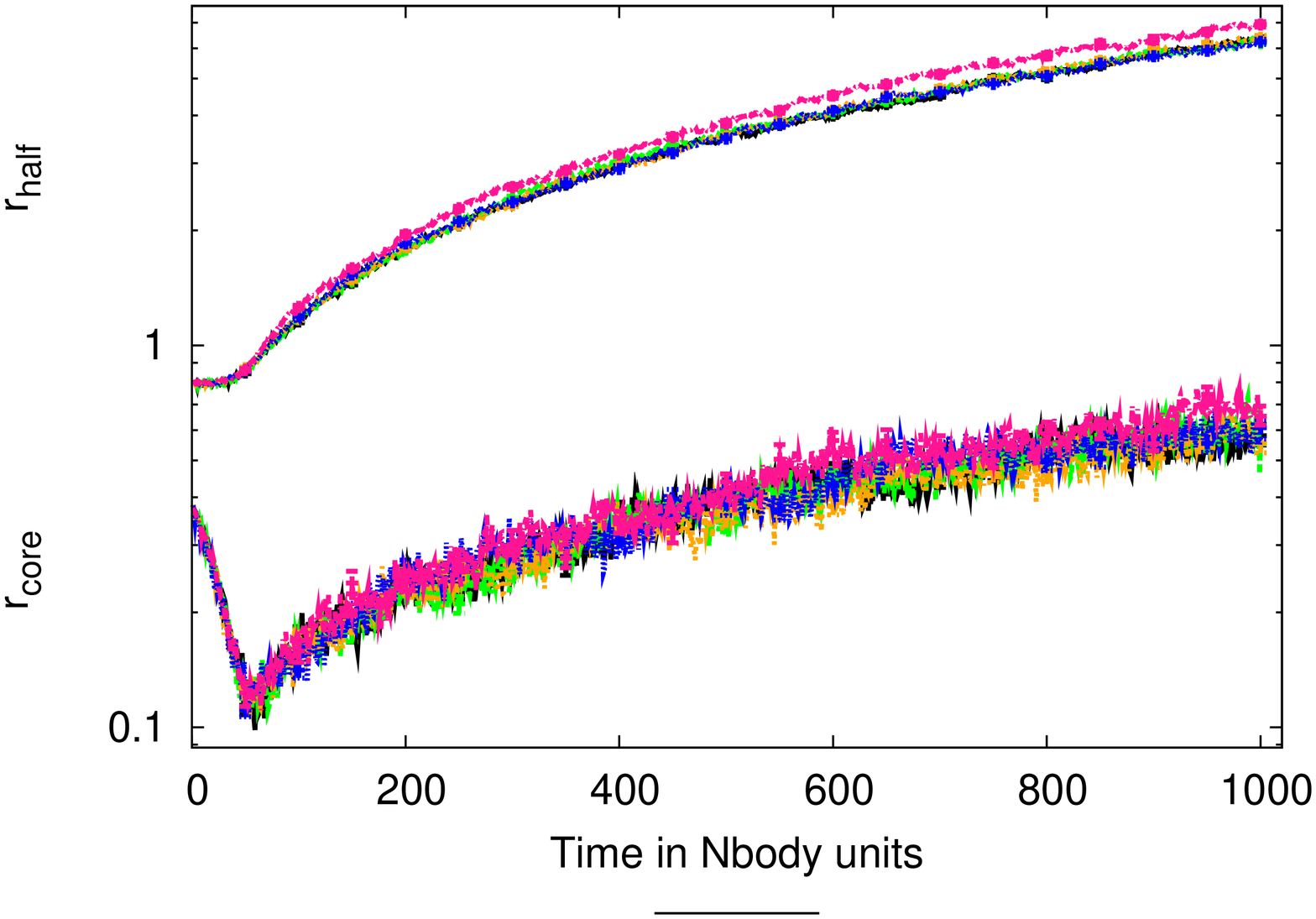} & 
   \includegraphics[angle=270,width=0.37\linewidth]{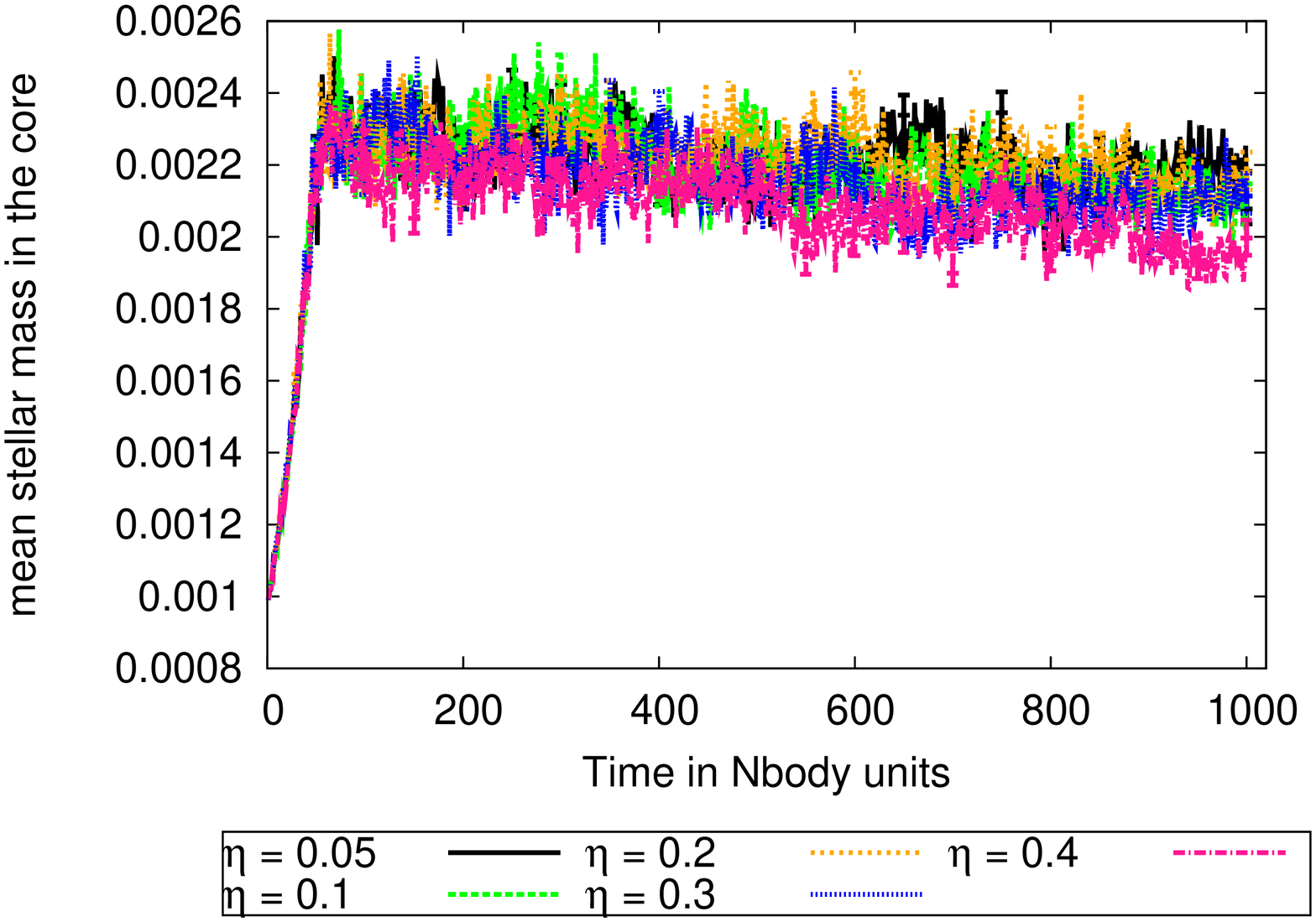} \\ 
   \includegraphics[angle=270,width=0.37\linewidth]{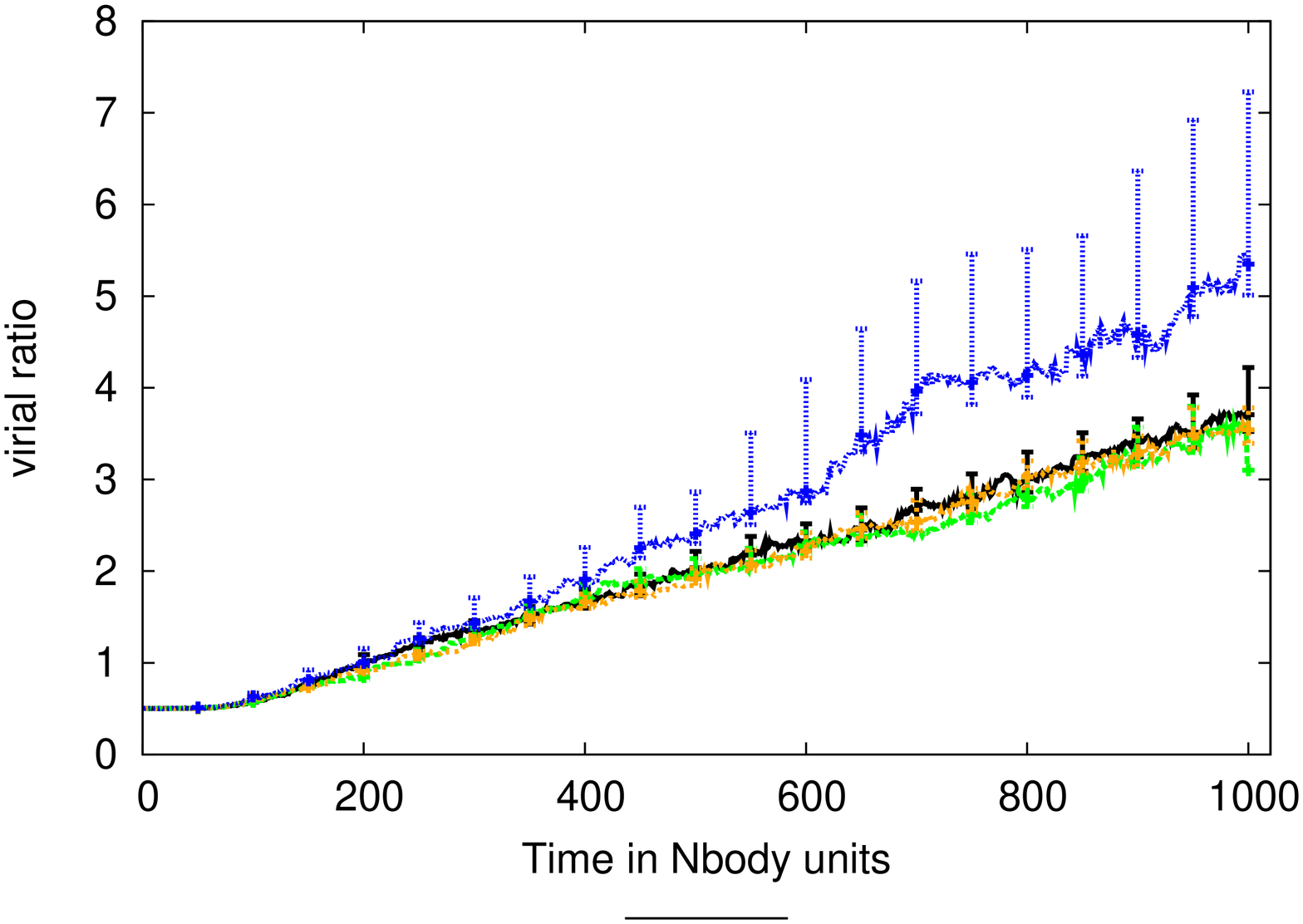} & 
   \includegraphics[angle=270,width=0.37\linewidth]{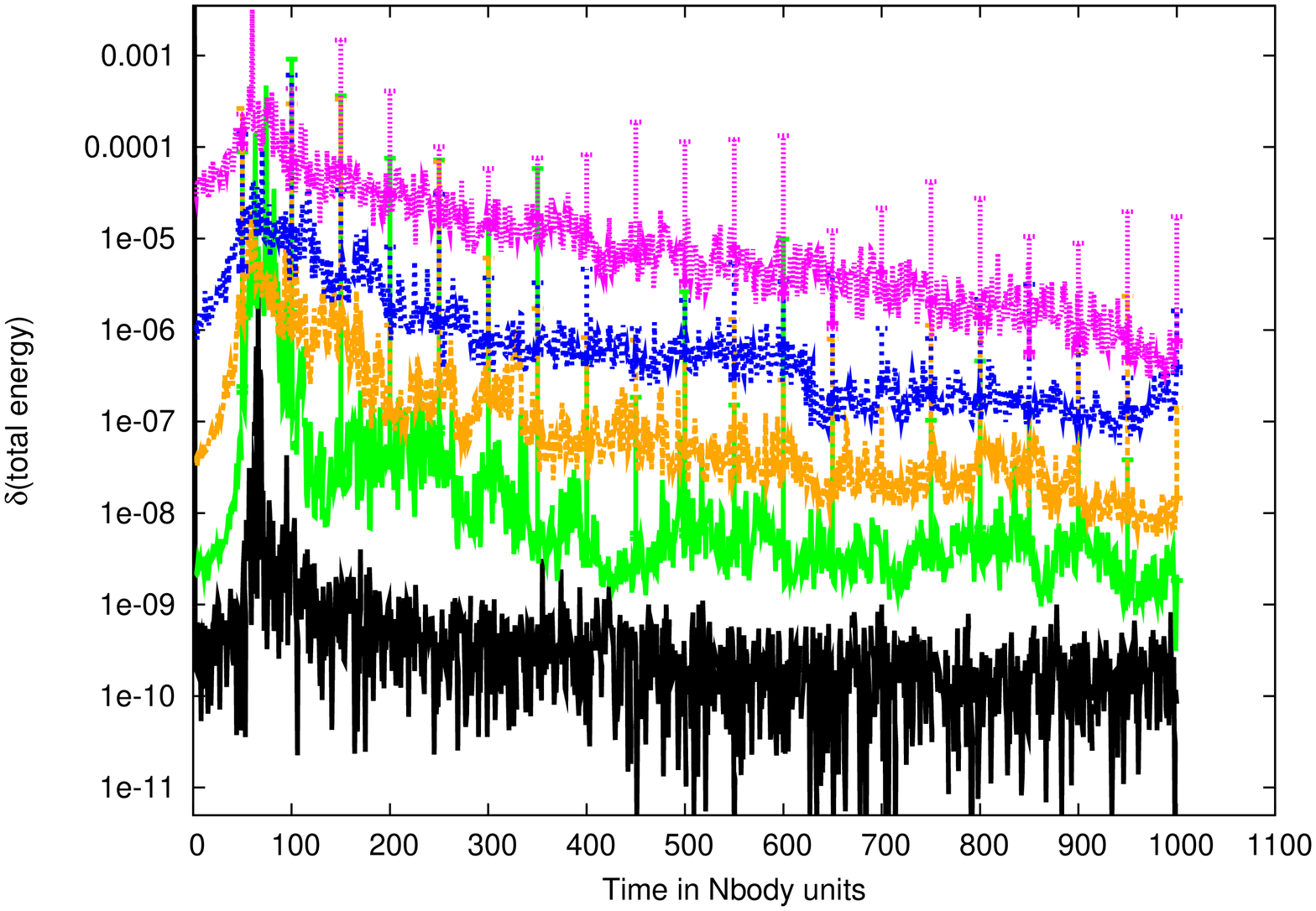} \\ 
  \end{tabular}
\end{center}

   \caption{Comparison of simulations using {\sc starlab} and various
   accuracy settings, all using PC1 and GPU.  The lines show the median values, the error bars
   give the uncertainty ranges from the 50 individual runs. Shown are
   the time evolutions of the core radius (top left, bottom lines),
   half-mass radius (top left, upper lines), the objects' mean mass in
   the core =  measure for mass segregation (top right), virial ratio
   (bottom left; $\eta$=0.4 is not shown due to very large deviations and error bars) and change in total energy during 1 $N$-body time unit 
   (bottom right).}

\label{fig:acc_struct_par}

\begin{center}
   \begin{tabular}{cc}
     \includegraphics[angle=270,width=0.37\linewidth]{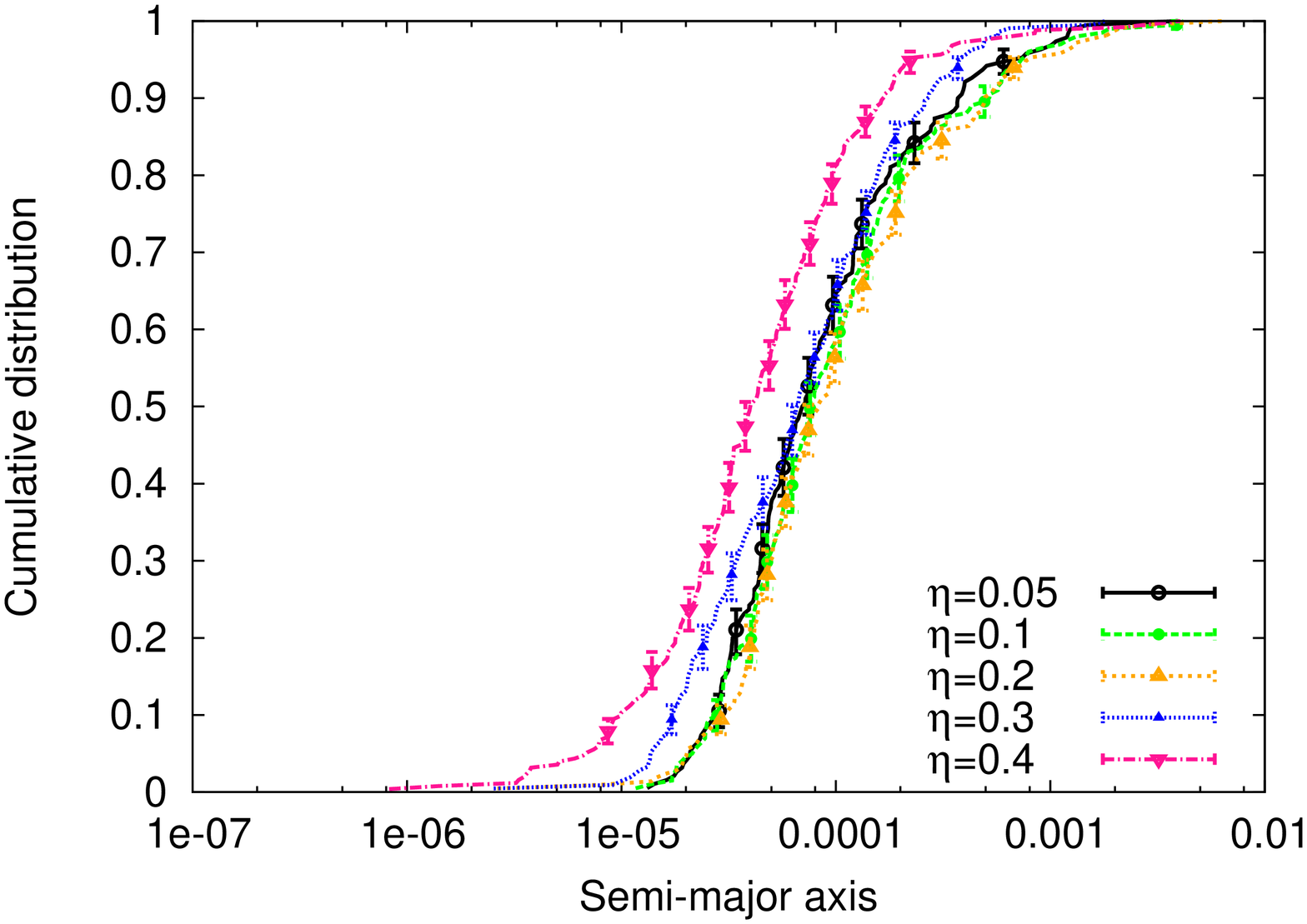} &
     \includegraphics[angle=270,width=0.37\linewidth]{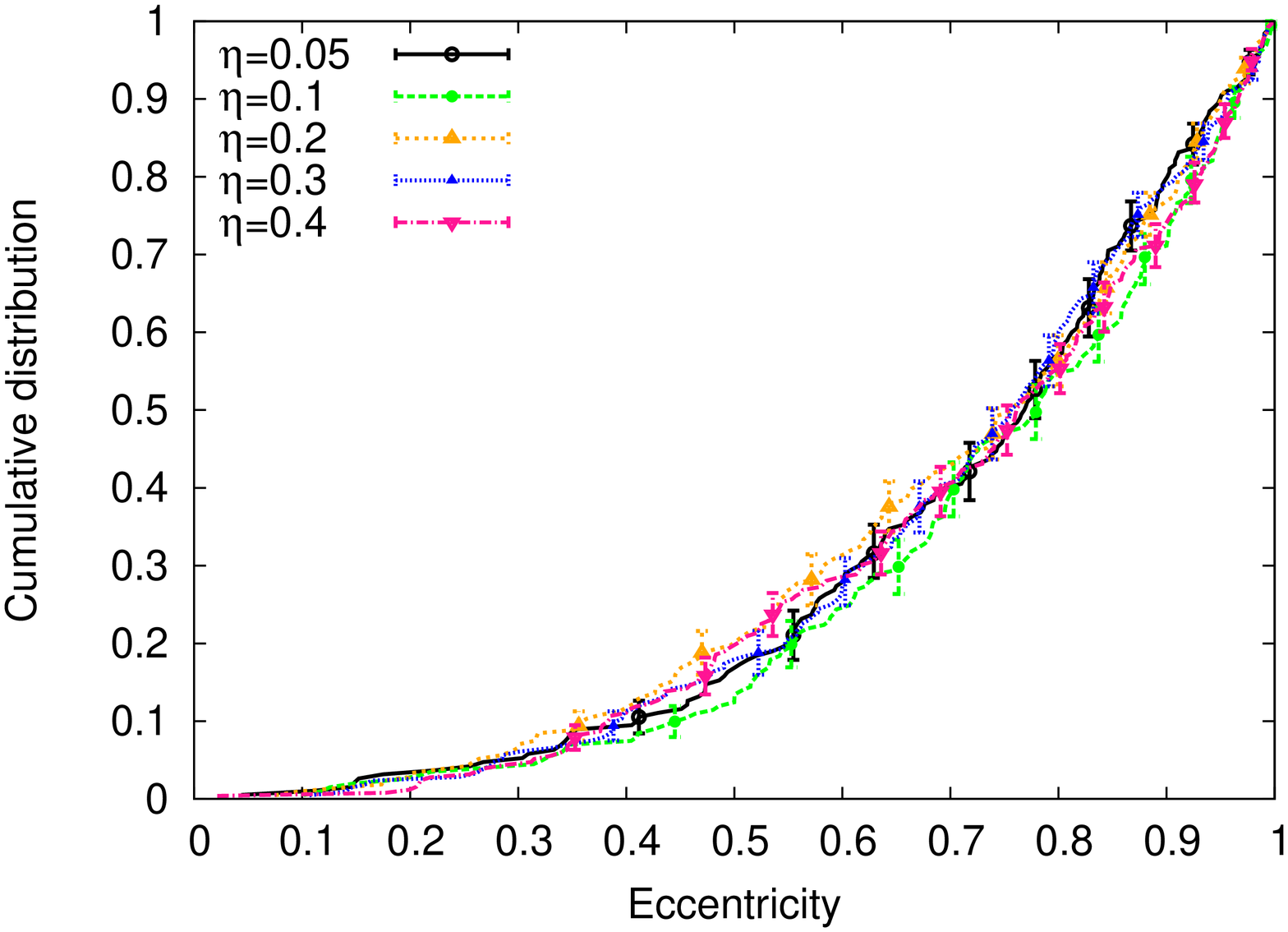} \\  
     \includegraphics[angle=270,width=0.37\linewidth]{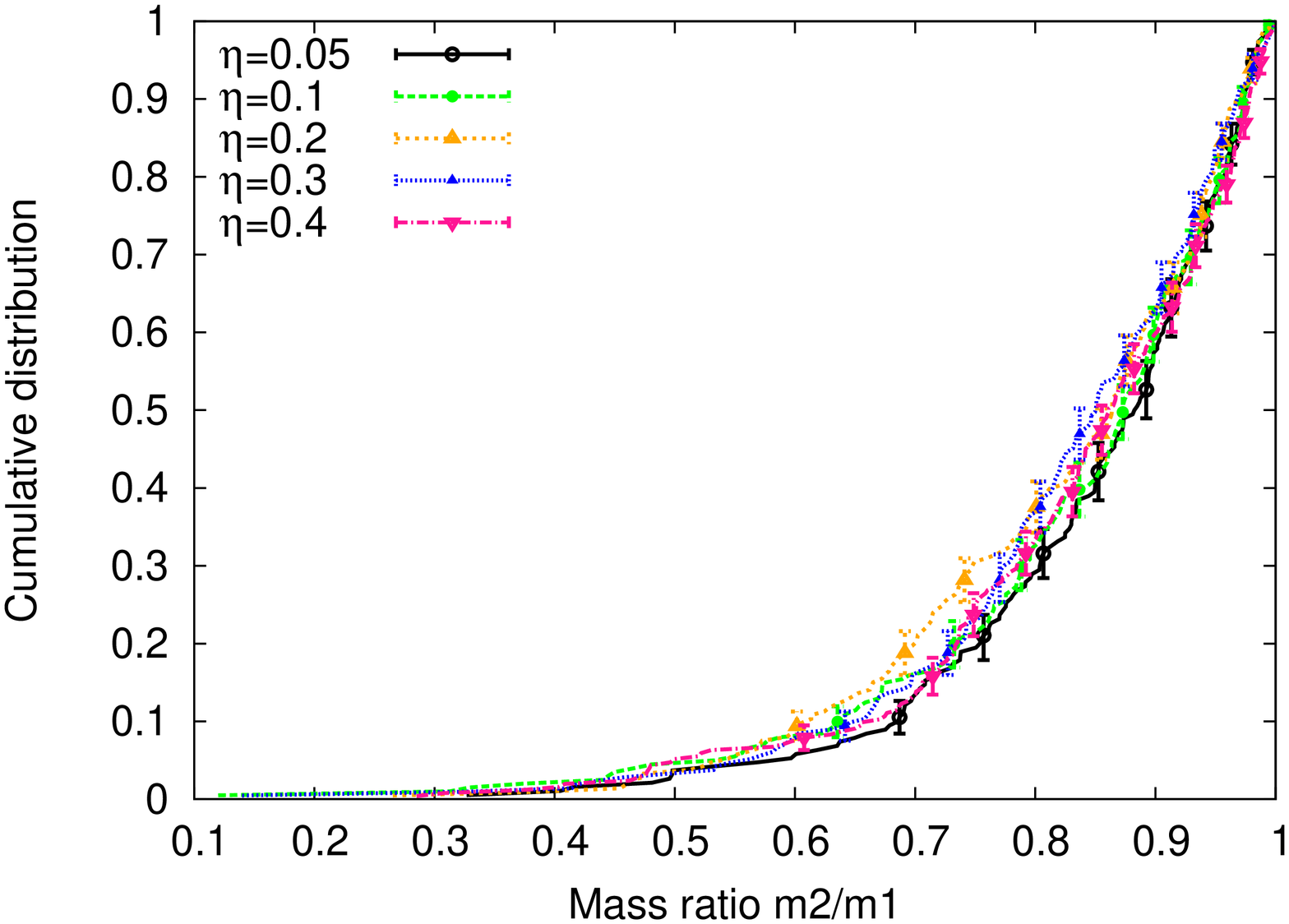} &
     \includegraphics[angle=270,width=0.37\linewidth]{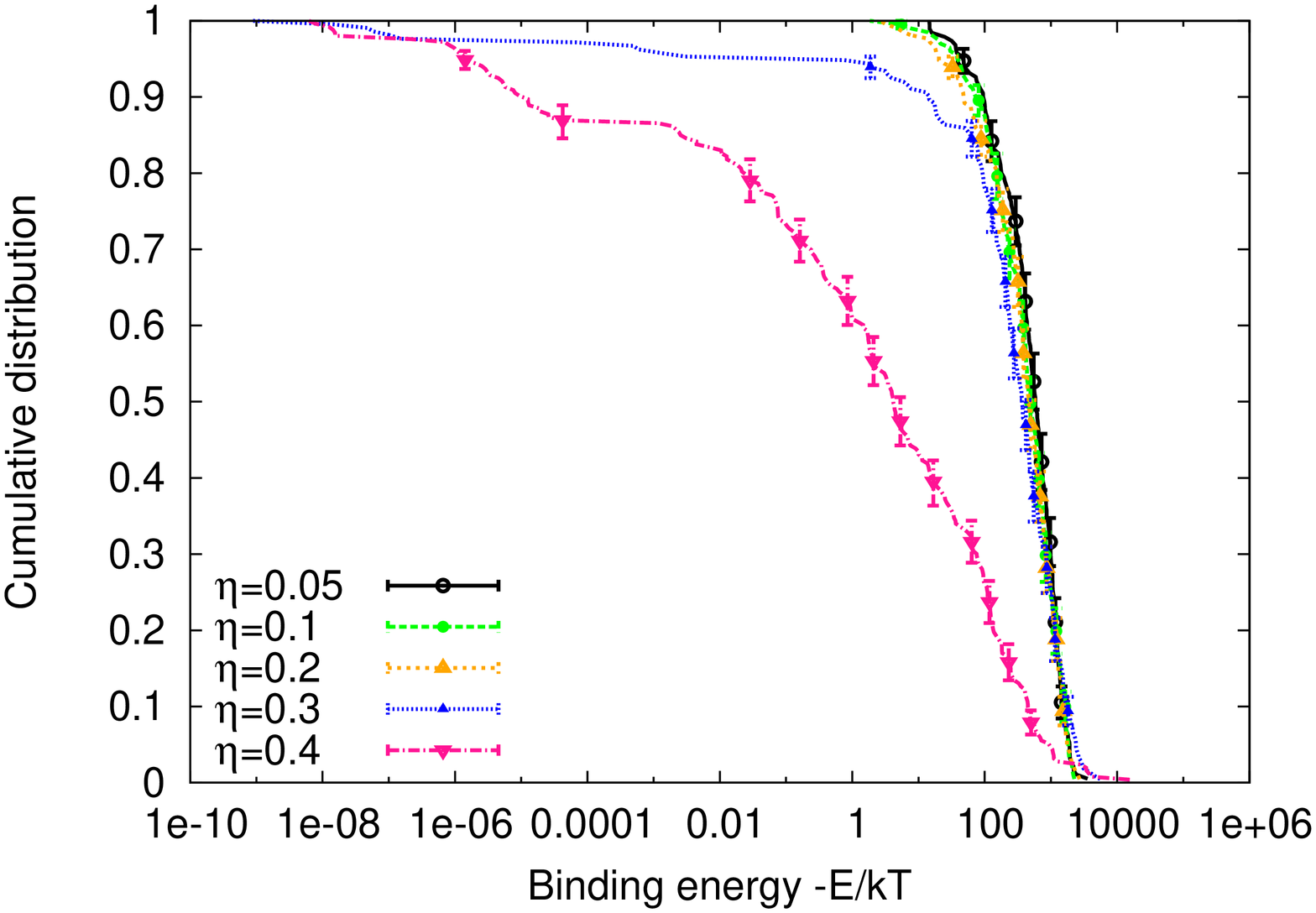} \\
   \end{tabular}

   \caption{Comparison of binary parameters from simulations using  {\sc
   starlab} and various accuracy settings after 1000 $N$-body time units
   (i.e., well after core collapse). All using PC1 and GPU. Shown are the cumulative
   distributions of the semi-major axis  (top left), the eccentricity
   (top right), mass ratio of secondary to primary  (bottom left) and
   the binding energy (bottom right). The lines show the data, the error
   bars give the uncertainty ranges from bootstrapping (at every 
   20$^{th}$ data point only, for clarity).}

   \label{fig:acc_bin_par}
\end{center}
\end{figure*}

The results are displayed in Figs. \ref{fig:acc_struct_par} and
\ref{fig:acc_bin_par}. 

In most panels, the simulations with the largest accuracy parameter (i.e., the
``worst'' accuracy) studied are clearly offset from the other simulations. The
other simulations seem to provide quite comparable results, except for the
change in total energy during 1 $N$-body time unit  (i.e., a measure for energy
conservation). The change in total energy reduces with lowering the accuracy
parameter $\eta$ both before and after core collapse, i.e. total energy becomes
better preserved when improving accuracy. 

These findings are confirmed by the quantitative measures from our tests: 

\begin{itemize}

\item (highly) significant: most results for the $\eta$=0.3 and $\eta$=0.4  runs
strongly deviate from the results using the standard settings, independent  of
used hardware and studied parameter. For the $\eta$=0.25 runs, the results depend
on the studied parameter: one part of the results follow the standard setting
results, while for the other parameters the results deviate from the standard
setting runs similarly to the $\eta$=0.3 run results (no clear correlation between
studied parameter and found deviation)

\item (highly) significant: the change in total energy during 1 $N$-body
time unit (i.e. our measure for energy conservation) is deviating from the 
standard settings for almost all runs

\item (highly) significant: additional individual deviations are found,
without a specific correlation with simulation settings, used hardware 
or studied parameter.

\end{itemize}

The strongly deviating results for $\eta$=0.3 and $\eta$=0.4 runs (and partially
the $\eta$=0.25 runs) represent strong discouragement for such ``inaccurate''
simulations.

The results for the energy conservation is expected. The better the simulation
accuracy the better even slight effects are traced. These slight effects, e.g. 
individual close encounters or binary evolution, have strong impact on the
conservation of the system's energy.

The number and level of additional deviations exceeds the expectations based on
``multiple  testing'', also without the $\eta$=0.3 and $\eta$=0.4 runs. The level
of accuracy changes the simulation results slightly but cumulatively, therefore
fundamental agreement for all studied parameters from simulations using different
accuracy settings is not assured.

\subsection{{\sc starlab} vs. {\sc nbody6}}

We have performed simulations on PC1 and PC3. For each PC we realised one set of
simulations with the installed GPUs and one set without hardware acceleration.
The {\sc starlab} simulations were performed with $\eta_{\rm starlab}$ = 0.1, while 
the {\sc nbody6} simulations were performed with $\eta_{\rm starlab} \approx$ 0.15 
($\eta_i = \sqrt{0.02}$ and $\eta_r = \sqrt{0.03}$). As shown in Sect. \ref{sec:acc},
these settings give comparable results.

\begin{figure*}
\begin{center}
  \begin{tabular}{cc}   
   \includegraphics[angle=270,width=0.37\linewidth]{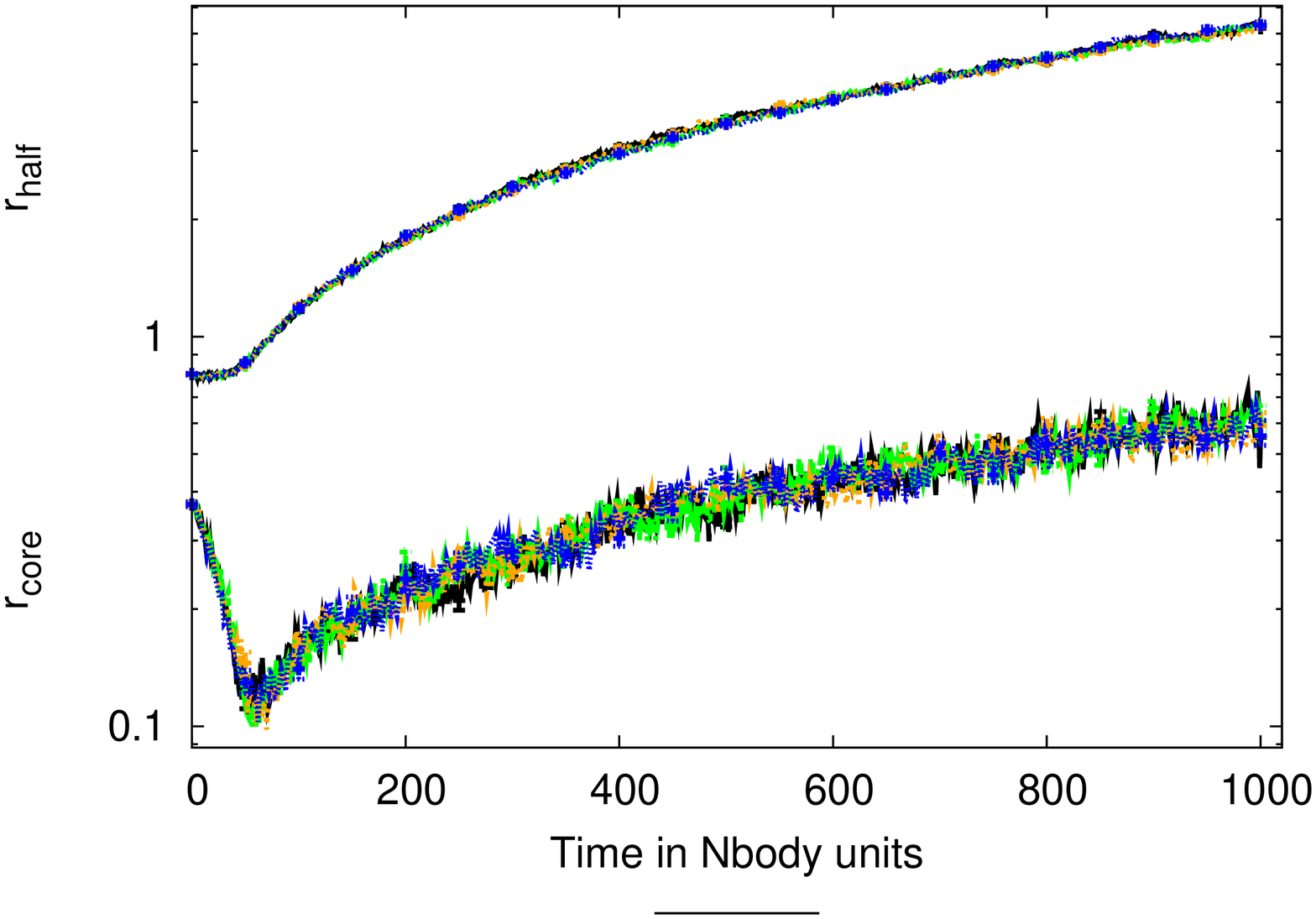} &
   \includegraphics[angle=270,width=0.37\linewidth]{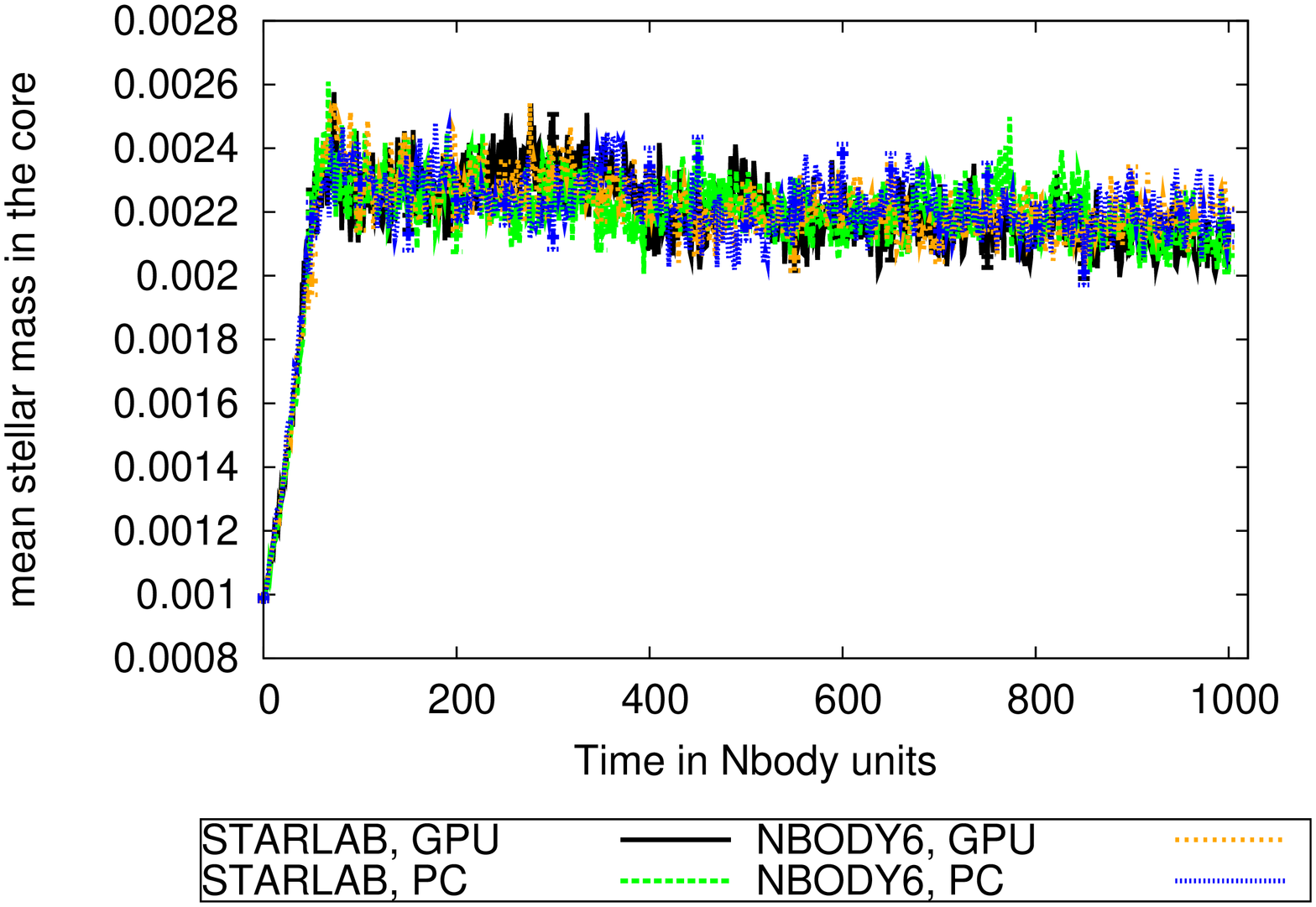} \\
   \includegraphics[angle=270,width=0.37\linewidth]{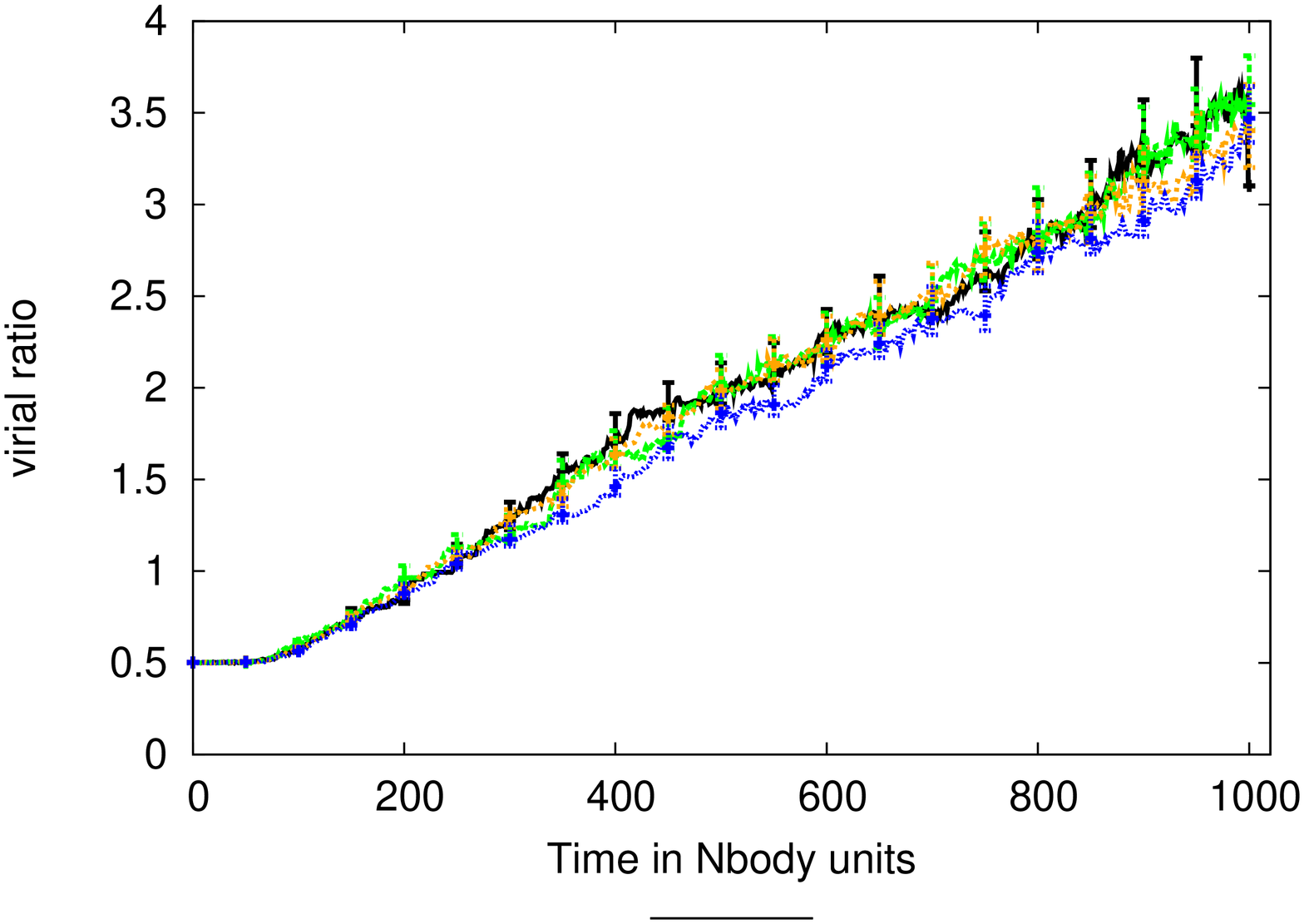} &
   \includegraphics[angle=270,width=0.37\linewidth]{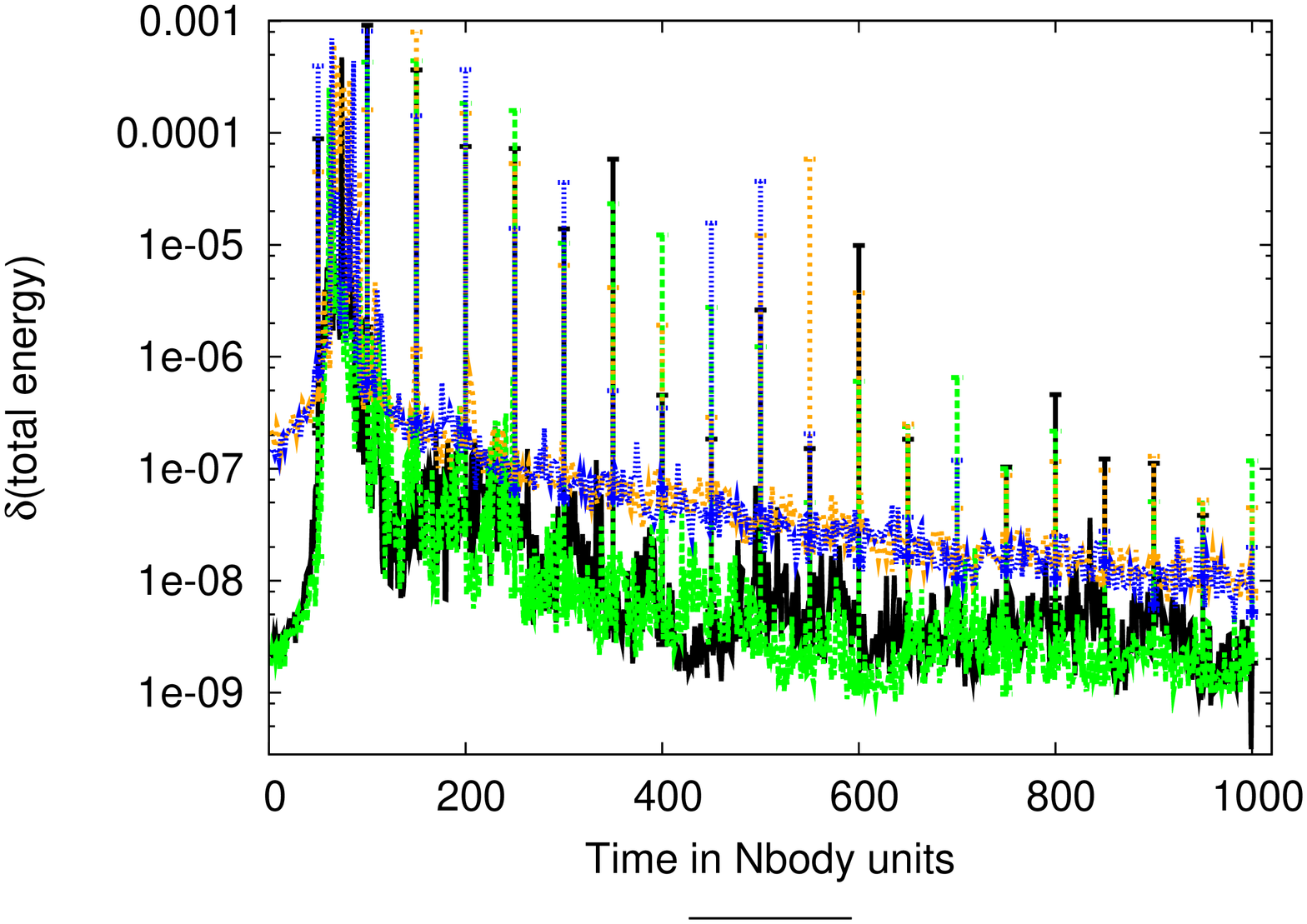} \\
  \end{tabular}
\end{center}

   \caption{Comparison of simulations using {\sc starlab} or {\sc
   nbody6}, with either GPU acceleration or no hardware acceleration. 
   The lines show the median values, the error bars give the uncertainty
   ranges from the 50 individual runs. Shown are the time evolutions of
   the core radius (top left, bottom lines), half-mass radius (top left,
   upper lines), the objects' mean mass in the core =  measure for mass
   segregation (top right), virial ratio (bottom left) and change in
   total energy during 1 $N$-body time unit (bottom right).}

\label{fig:nbsl_struct_par}

\begin{center}
   \begin{tabular}{cc}
     \includegraphics[angle=270,width=0.37\linewidth]{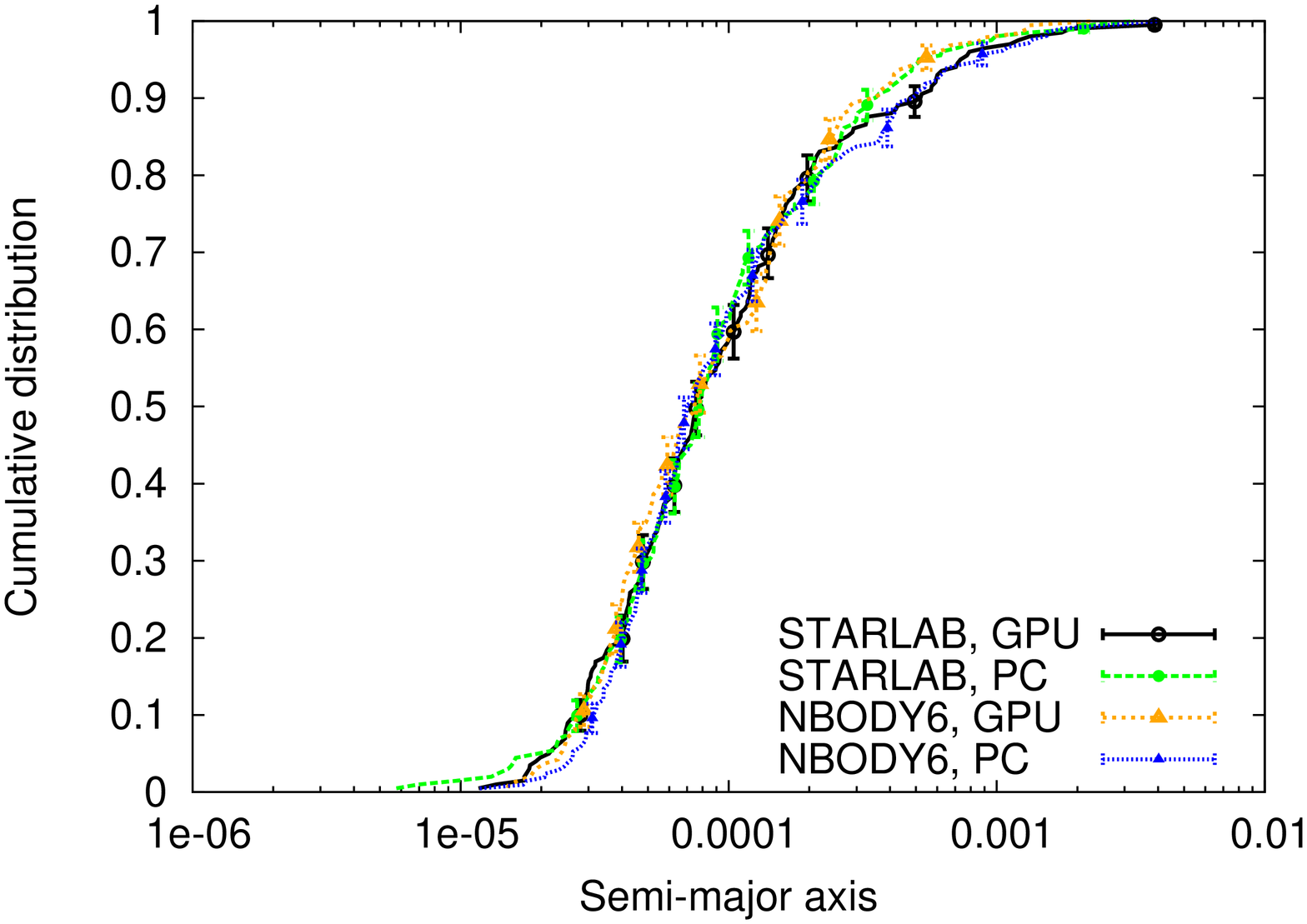} &
     \includegraphics[angle=270,width=0.37\linewidth]{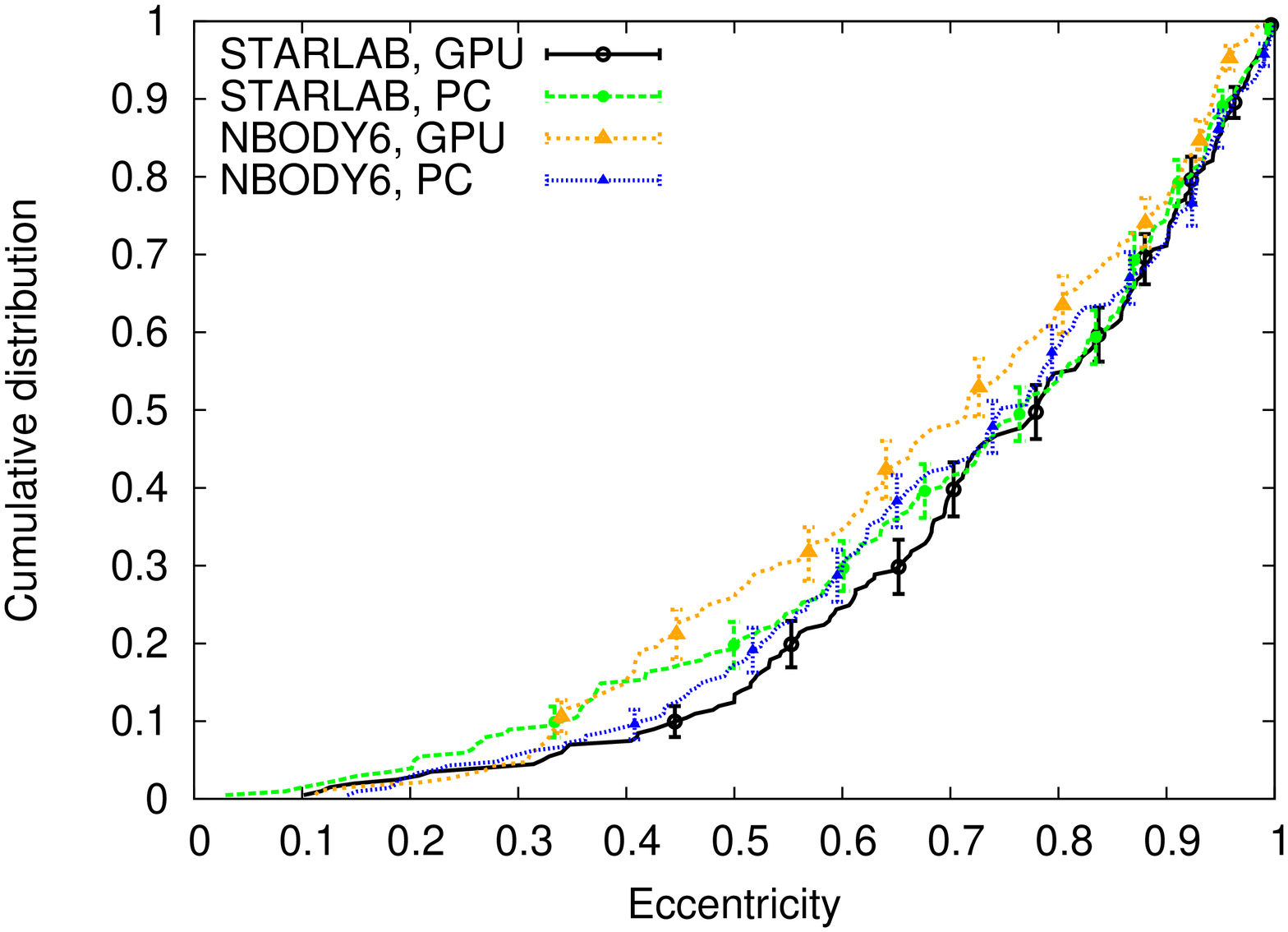} \\
     \includegraphics[angle=270,width=0.37\linewidth]{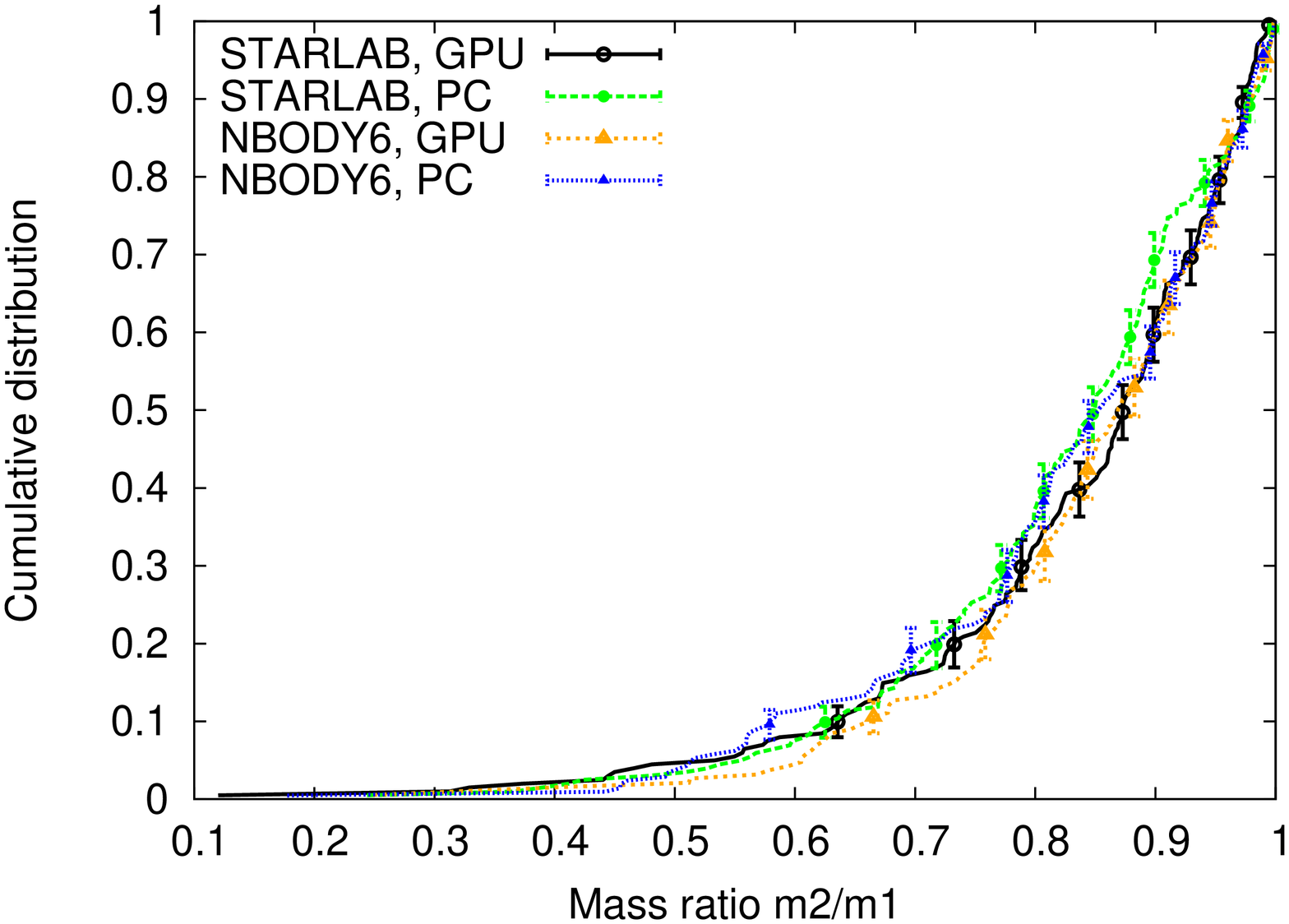} &
     \includegraphics[angle=270,width=0.37\linewidth]{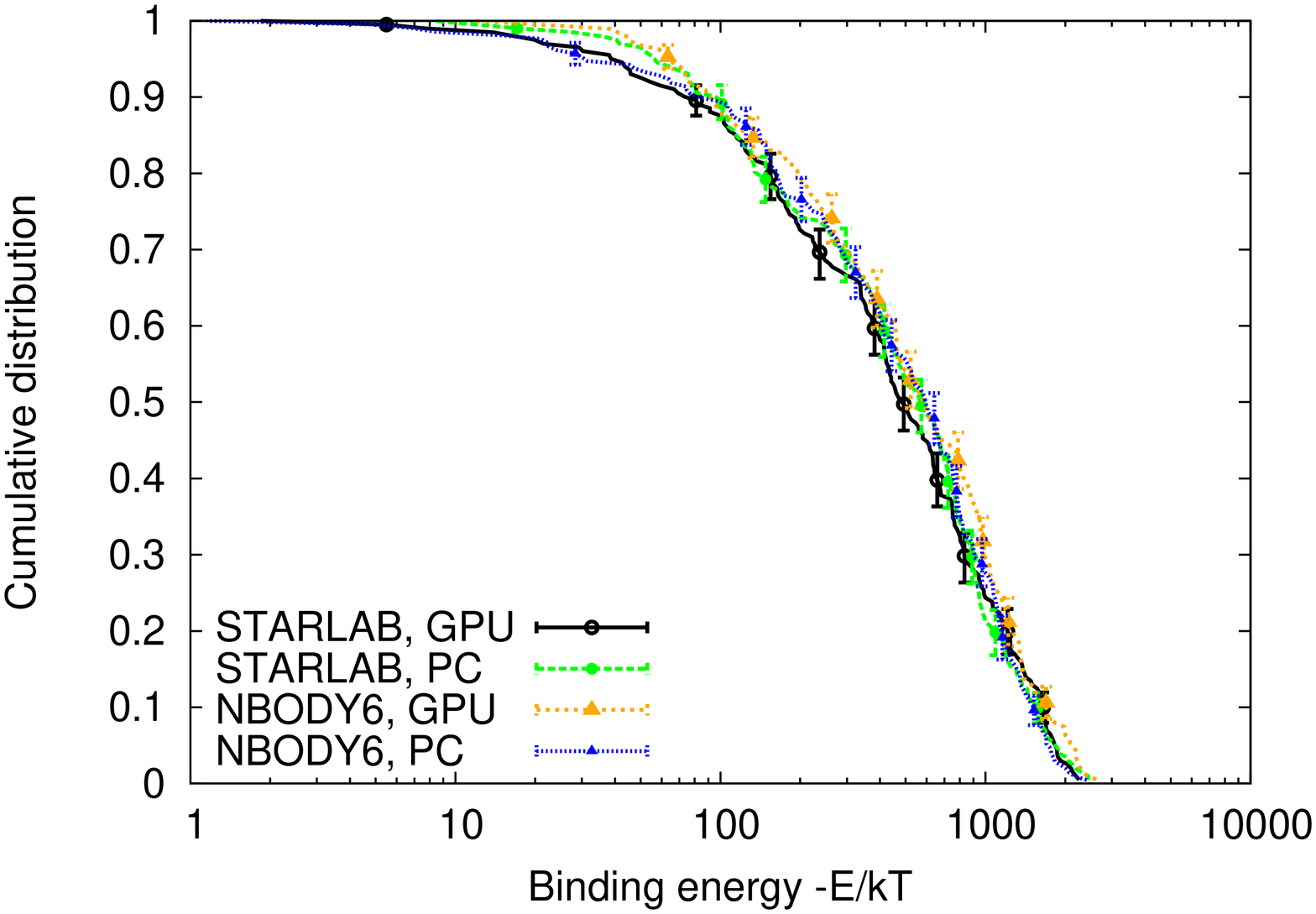} \\
   \end{tabular}

   \caption{Comparison of binary parameters from simulations using  {\sc
   starlab} or {\sc nbody6}, with either GPU acceleration or no hardware
   acceleration, after 1000 $N$-body time units (i.e., well after core
   collapse). Shown are the cumulative distributions of the semi-major
   axis  (top left), the eccentricity (top right), mass ratio of
   secondary to primary  (bottom left) and the binding energy (bottom
   right). The lines show the data, the error
   bars give the uncertainty ranges from bootstrapping (at every 
   20$^{th}$ data point only, for clarity).}

   \label{fig:nbsl_bin_par}
\end{center}
\end{figure*}

The results are shown in Figs. \ref{fig:nbsl_struct_par} and
\ref{fig:nbsl_bin_par}. They indicate potential deviations in kinetic energy for
the {\sc nbody6} runs using PC3, in the change of total energy (during 1
$N$-body time unit) between {\sc nbody6} and  {\sc starlab} runs, and in binary
eccentricity for the GPU runs using  either {\sc nbody6} or {\sc starlab}.

Our quantitative tests confirm the deviations for the change in total energy
during 1 $N$-body time unit: It is bigger (i.e. energy conservation is worse) for
{\sc nbody6} as compared to {\sc starlab}. This agrees with our earlier finding
for {\sc nbody4} vs. {\sc starlab} in Paper I.

The remaining expectations, based on visual examination of Figs.
\ref{fig:nbsl_struct_par} and \ref{fig:nbsl_bin_par}, are not confirmed.
Our statistical tests indicate deviations in 

\begin{itemize}

\item significant: the time evolution of r$_{max}$ 
(i.e., the distance of the furthest star from the cluster center) 
when comparing simulations made with {\sc starlab} and {\sc nbody6}, 
both without hardware acceleration

\item weakly significant: the binary mass ratio distribution for \#3.2 PC1 $\Leftrightarrow$ PC3

\item weakly significant: the distribution the kinetic
energy of stars escaping the cluster for \#3.1 PC1/GPU $\Leftrightarrow$ PC3/GPU

\end{itemize}

These results are partially inconsistent with our previous study of {\sc starlab} 
vs {\sc nbody4} in Paper I. In Paper I no significant deviations in r$_{max}$
were  found, the binary mass ratio distributions were different (the probability
of  both being drawn from the same parent distribution was determined to be
14.9\%,  slightly larger than required for a statistically slightly significant
deviation) and  for the kinetic energy of stars escaping the cluster no
significant deviations was found. The detected differences in the present work are
expected to originate from code differences between {\sc nbody4} and {\sc nbody6}
(code enhancements, such as the AC neighbour scheme)  or {\sc starlab} (version
4.4.2) and {\sc starlab} (version 4.4.4) (which included bug fixes). 

The results for the time evolution of r$_{max}$ and the binary mass ratio distributions 
are based on processes with strong statistical effects as these quantities are affected by 
rare events, e.g. direct two-body interaction processes. In general, the number and level
of deviations agrees with the expectations from ``multiple testing''. These results allow
reliable comparisons of {\sc nbody6} and {\sc starlab} simulations, based on comparable 
input parameters and independent of the used computer hardware.

\section{Summary of results \& conclusions}
\label{sec:summary}

All of the quantitative test results are shown in Tables
\ref{tab:res_boot_struct} - \ref{tab:kstest_esc}. They are
based on 50 input models and their consistently analysed output.

From these tables the following picture emerges

\begin{itemize}

\item in general, the results obtained from $N$-body simulations in various
configurations are well comparable (with some exceptions which will be discussed
as follows), independent of the computer  hardware, the accelerator type and
$N$-body code used.

\item results with too coarse accuracy (for our tests, $\eta_{STARLAB}$ = 0.3 and $\eta_{STARLAB}$ = 0.4, partially
$\eta_{STARLAB}$ = 0.25) give strongly significantly deviating results in a variety of quantities, hence should be avoided. 

\item the quantity which shows the highest sensitivity to the configuration
used is the energy conservation (here described by the change of total
cluster energy during one $N$-body time unit). While it seems to be
unaffected by the hardware used, it changes strongly with the accuracy
parameter $\eta$, and between {\sc nbody6} and {\sc starlab}. The
behaviour with accuracy parameter is expected and desired. The differences
between {\sc nbody6} and {\sc starlab} are comparable to our earlier
finding when comparing {\sc nbody4} and {\sc starlab} in
Paper I.

\item the significant deviation found for some configurations in the
temporal evolution of r$_{max}$ (distance of furthest star from cluster
center) and the mass ratio distribution of binaries could be stochastic
effects, as these quantities are affected by rare events (e.g. close two-body 
encounters and binary evolution). In our test 
setting, these quantities deviate statistically significant, though larger-scale
simulations are required to fundamentally check the statistical significance.

\item the remaining significant deviations can be real or a result of
``multiple testing''. The number of performed tests and of observed deviations
is consistent  with resulting from ``multiple testing'' for the study of
hardware (\#1.1-1.4 PC1 $\Leftrightarrow$ PC2, utilising or relinquishing the available 
accelerator hardware) and of the used $N$-body code
(\#3.1-3.3 {\sc starlab} $\Leftrightarrow$ {\sc nbody6}). For the study of accuracy
settings the number and strength of deviations is higher than expected from
``multiple testing'' for the energy-related cluster properties and the binary
properties. 

\end{itemize}

To conclude: In general, direct N-body simulations are assumed to be correct in a statistical sense rather then on the
level of individual trajectories. Except for few well-understood cases (such as the results for various accuracy
settings and the energy conservation), $N$-body simulations using different hardware or $N$-body codes give very
comparable results, confirming the reliability in a statistical sense.

{\bf Acknowledgements} 

P.A. acknowledges funding by the European Union (Marie Curie EIF grant
MEIF-CT-2006-041108), the National Natural Science Foundation of
China (NSFC, grant number 11073001) and wishes to thank N. Bissantz for helpful discussions. 
H.B. acknowledges support from the Australian Research
Council through Future Fellowship grant FT0991052 and thanks K. Nitadori for
providing the GPU module for {\sc nbody6}. SPZ thanks the Netherlands Research Council
NWO (VICI [\#639.073.803]) and the Netherlands Research School for
Astronomy (NOVA). Support for Program number HST-HF-51289.01-A was provided by NASA through a Hubble Fellowship grant 
from the Space Telescope Science Institute, which is operated by the Association of Universities for Research in Astronomy, 
Incorporated, under NASA contract NAS5-26555.

\bibliographystyle{aa}
\bibliography{NB_SL_Comp}

\appendix

\section{Graphical display of simulation results}
\label{app:graphics}

\section{Analysis results}
\label{app:tables}

\begin{table*}

\caption{Bootstrap results for structural cluster properties. Given are
the fractions (in \%) of the test cluster distributions (which represent 
the case that they are drawn from the same parent distribution) more deviating than the
main simulations, i.e. the smaller this number the less alike the
distributions are. The ``standard'' setup uses 
PC1/GPU and an accuracy setting of $\eta$=0.1. Quantities are:
r$_{\rm core}$ = cluster core radius, 
r$_{\rm half}$ = cluster half-mass radius, 
r$_{\rm max}$ = distance of furthest star from cluster center,
$<$mass$>_{\rm core}$ =  mean mass of objects in the core (i.e. a measure for
mass segregation), 
King W$_0$ = dimensionless potential W$_0$ of a King profile (fitted by the {\sc hsys$\_$stats} 
analysis task of {\sc starlab}), 
$\arrowvert$ density c. $\arrowvert$ = absolute distance of cluster density 
center from origin. 
For each parameter the probabilities from the $\Delta_{12}$ and $\Gamma_{12}$
distributions (see Sect. \ref{sec:methods.timeseries}) are given. If no $\eta$ value is
shown, $\eta$=0.1 is used. For a more detailed description of the setups see
text. Notes:  $^e$ denotes data sets with ``extreme'' uncertainties, i.e.
``extreme run-to-run'' scatter (applies only to the $\eta$=0.4 data sets).}

\begin{center}
\begin{tabular}{l l l c  c c  c c c}
\hline

\#ID & setup1 & setup2                & r$_{\rm core}$ & r$_{\rm half}$ & r$_{\rm max}$ & King W$_0$ & $\arrowvert$ density c. $\arrowvert$ & $<$mass$>_{\rm core}$\\
       &      &                       & $\Delta_{12}$  $\Gamma_{12}$ & $\Delta_{12}$  $\Gamma_{12}$ & $\Delta_{12}$  $\Gamma_{12}$ & $\Delta_{12}$  $\Gamma_{12}$ & $\Delta_{12}$  $\Gamma_{12}$ & $\Delta_{12}$  $\Gamma_{12}$ \\
\hline
\multicolumn{7}{|l|}{\bf{Tests using different PCs and accelerator hardware}} \\
\scriptsize{1.1} & \scriptsize{PC1/GPU} & \scriptsize{PC1} & \scriptsize{69.84 90.39} & \scriptsize{40.34 69.53} & \scriptsize{32.25 40.35} & \scriptsize{29.99 62.47} & \scriptsize{65.88 81.42} & \scriptsize{80.62 31.04} \\
\scriptsize{1.2} & \scriptsize{PC1/GPU} & \scriptsize{PC2/GRAPE} & \scriptsize{53.04 21.16} & \scriptsize{91.99 75.21} & \scriptsize{72.32 88.50} & \scriptsize{32.04 22.54} & \scriptsize{38.78 47.13} & \scriptsize{35.55 39.60} \\
\scriptsize{1.3} & \scriptsize{PC2/GRAPE} & \scriptsize{PC2} & \scriptsize{31.78 42.99} & \scriptsize{35.33 69.37} & \scriptsize{18.52 21.18} & \scriptsize{{\bf 5.76} 33.24} & \scriptsize{67.62 74.21} & \scriptsize{48.73 72.74} \\
\scriptsize{1.4} & \scriptsize{PC1} & \scriptsize{PC2} & \scriptsize{99.76 99.94} & \scriptsize{99.89 99.94} & \scriptsize{99.94 99.94} & \scriptsize{99.92 99.94} & \scriptsize{99.91 99.94} & \scriptsize{99.89 99.94} \\
\hline
\hline
\multicolumn{7}{|l|}{\bf{Tests using different accuracy settings $\eta$}} \\
\scriptsize{2.1} & \scriptsize{standard} & \scriptsize{PC1/GPU} $\eta$ \scriptsize{=0.05} & \scriptsize{90.53 78.08} & \scriptsize{33.26 47.85} & \scriptsize{45.83 54.34} & \scriptsize{80.62 56.33} & \scriptsize{16.17 16.73} & \scriptsize{63.13 56.13} \\
\scriptsize{2.2} & \scriptsize{standard} & \scriptsize{PC1/GPU} $\eta$ \scriptsize{=0.2} & \scriptsize{24.23 62.85} & \scriptsize{84.23 59.07} & \scriptsize{59.08 69.36} & \scriptsize{22.56 37.74} & \scriptsize{39.55 46.59} & \scriptsize{39.16 30.37} \\
\scriptsize{2.3} & \scriptsize{standard} & \scriptsize{PC1/GPU} $\eta$ \scriptsize{=0.3} & \scriptsize{16.76 11.23} & \scriptsize{51.91 22.42} & \scriptsize{{\bf  6.67  8.25}} & \scriptsize{{\bf  3.68  3.74}} & \scriptsize{48.51 61.55} & \scriptsize{{\bf  5.93  9.22}} \\
\scriptsize{2.4} & \scriptsize{standard} & \scriptsize{PC1/GPU} $\eta$ \scriptsize{=0.4} & \scriptsize{{\bf 0.10  2.24}} & \scriptsize{{\bf 0.76  0.79}} & \scriptsize{{\bf 0.00  0.00} $^e$} & \scriptsize{{\bf 1.48  2.68}} & \scriptsize{{\bf 0.01  0.01} $^e$} & \scriptsize{{\bf 0.14  1.84}} \\
\hline
\scriptsize{2.5} & \scriptsize{PC1/GPU} $\eta$ \scriptsize{=0.05} & \scriptsize{PC1} $\eta$ \scriptsize{=0.05} & \scriptsize{63.07 31.75} & \scriptsize{12.97 17.48} & \scriptsize{84.45 73.63} & \scriptsize{90.32 33.38} & \scriptsize{22.55 16.88} & \scriptsize{71.13 54.56} \\
\scriptsize{2.6} & \scriptsize{PC1/GPU} $\eta$ \scriptsize{=0.3} & \scriptsize{PC1} $\eta$ \scriptsize{=0.3} & \scriptsize{40.35 62.27} & \scriptsize{78.80 90.92} & \scriptsize{{\bf 2.92  3.19}} & \scriptsize{73.41 58.60} & \scriptsize{23.97 27.10} & \scriptsize{84.90 73.96} \\
\scriptsize{2.7} & \scriptsize{PC1/GPU} $\eta$ \scriptsize{=0.4} & \scriptsize{PC1} $\eta$ \scriptsize{=0.4} & \scriptsize{{\bf  5.76} 17.15} & \scriptsize{30.84 54.66} & \scriptsize{{\bf 4.56  4.55} $^e$} & \scriptsize{83.53 27.20} & \scriptsize{12.03 14.27 $^e$} & \scriptsize{{\bf 2.26  5.77}} \\
\hline
\scriptsize{2.8} & \scriptsize{PC1} $\eta$ \scriptsize{=0.1} & \scriptsize{PC1} $\eta$ \scriptsize{=0.025} & \scriptsize{38.52 89.72} & \scriptsize{62.60 85.86} & \scriptsize{{\bf 6.54  4.90}} & \scriptsize{17.27 63.72} & \scriptsize{85.57 72.98} & \scriptsize{33.01 58.43} \\
\scriptsize{2.9} & \scriptsize{PC1} $\eta$ \scriptsize{=0.1} & \scriptsize{PC1} $\eta$ \scriptsize{=0.05} & \scriptsize{83.51 72.57} & \scriptsize{16.69 29.78} & \scriptsize{65.07 53.41} & \scriptsize{35.25 52.36} & \scriptsize{81.28 74.73} & \scriptsize{53.60 76.91} \\
\scriptsize{2.10} & \scriptsize{PC1} $\eta$ \scriptsize{=0.1} & \scriptsize{PC1} $\eta$ \scriptsize{=0.3} & \scriptsize{{\bf 7.73} 65.07} & \scriptsize{63.79 88.17} & \scriptsize{{\bf 0.18  0.25}} & \scriptsize{10.86 34.68} & \scriptsize{{\bf 1.58  1.84}} & \scriptsize{{\bf 2.82  4.32}} \\
\scriptsize{2.11} & \scriptsize{PC1} $\eta$ \scriptsize{=0.1} & \scriptsize{PC1} $\eta$ \scriptsize{=0.4} & \scriptsize{{\bf 6.12} 41.94} & \scriptsize{{\bf 0.78  0.79}} & \scriptsize{{\bf 0.45  0.58} $^e$} & \scriptsize{{\bf 2.54  7.09}} & \scriptsize{{\bf 7.90} 10.02 $^e$} & \scriptsize{{\bf 1.75  6.19}} \\
\hline
\scriptsize{2.12} & \scriptsize{PC2/GRAPE} $\eta$ \scriptsize{=0.1} & \scriptsize{PC2/GRAPE} $\eta$ \scriptsize{=0.05} & \scriptsize{12.53 {\bf  4.09}} & \scriptsize{79.69 85.86} & \scriptsize{41.02 41.70} & \scriptsize{{\bf  5.48  5.79}} & \scriptsize{84.51 91.26} & \scriptsize{61.87 15.14} \\
\scriptsize{2.13} & \scriptsize{PC2/GRAPE} $\eta$ \scriptsize{=0.1} & \scriptsize{PC2/GRAPE} $\eta$ \scriptsize{=0.2} & \scriptsize{60.32 70.07} & \scriptsize{64.11 91.51} & \scriptsize{45.93 62.23} & \scriptsize{31.19 55.00} & \scriptsize{82.72 92.35} & \scriptsize{40.92 91.06} \\
\scriptsize{2.14} & \scriptsize{PC2/GRAPE} $\eta$ \scriptsize{=0.1} & \scriptsize{PC2/GRAPE} $\eta$ \scriptsize{=0.25} & \scriptsize{36.41 22.06} & \scriptsize{52.11 37.57} & \scriptsize{{\bf 2.40  2.86}} & \scriptsize{12.73 25.07} & \scriptsize{77.08 86.30} & \scriptsize{34.87 22.32} \\
\scriptsize{2.15} & \scriptsize{PC2/GRAPE} $\eta$ \scriptsize{=0.1} & \scriptsize{PC2/GRAPE} $\eta$ \scriptsize{=0.3} & \scriptsize{98.73 27.86} & \scriptsize{82.25 91.92} & \scriptsize{{\bf 0.05  0.09}} & \scriptsize{84.11 54.87} & \scriptsize{{\bf 5.04  6.38}} & \scriptsize{29.73 58.32} \\
\hline
\scriptsize{2.16} & \scriptsize{PC2} $\eta$ \scriptsize{=0.1} & \scriptsize{PC2} $\eta$ \scriptsize{=0.025} & \scriptsize{35.15 88.64} & \scriptsize{64.30 84.28} & \scriptsize{{\bf  3.94  2.72}} & \scriptsize{13.60 59.63} & \scriptsize{77.85 51.25} & \scriptsize{28.70 49.31} \\
\scriptsize{2.17} & \scriptsize{PC2} $\eta$ \scriptsize{=0.1} & \scriptsize{PC2} $\eta$ \scriptsize{=0.05} & \scriptsize{83.47 72.21} & \scriptsize{16.73 29.86} & \scriptsize{65.07 53.41} & \scriptsize{35.19 52.63} & \scriptsize{81.35 74.71} & \scriptsize{53.64 77.51} \\
\scriptsize{2.18} & \scriptsize{PC2} $\eta$ \scriptsize{=0.1} & \scriptsize{PC2} $\eta$ \scriptsize{=0.2} & \scriptsize{36.16 76.20} & \scriptsize{43.41 50.17} & \scriptsize{65.45 87.24} & \scriptsize{41.31 61.96} & \scriptsize{88.75 94.54} & \scriptsize{89.62 55.47} \\
\scriptsize{2.20} & \scriptsize{PC2} $\eta$ \scriptsize{=0.1} & \scriptsize{PC2} $\eta$ \scriptsize{=0.25} & \scriptsize{41.90 81.57} & \scriptsize{11.63 17.46} & \scriptsize{20.53 17.35} & \scriptsize{10.52 34.10} & \scriptsize{72.97 27.49} & \scriptsize{83.06 46.80} \\
\scriptsize{2.21} & \scriptsize{PC2} $\eta$ \scriptsize{=0.1} & \scriptsize{PC2} $\eta$ \scriptsize{=0.3} & \scriptsize{{\bf  7.65} 64.17} & \scriptsize{63.92 88.13} & \scriptsize{{\bf  0.18  0.25}} & \scriptsize{10.53 33.50} & \scriptsize{{\bf  1.58  1.84}} & \scriptsize{{\bf  2.82  4.43}} \\
\scriptsize{2.22} & \scriptsize{PC2} $\eta$ \scriptsize{=0.1} & \scriptsize{PC2} $\eta$ \scriptsize{=0.4} & \scriptsize{{\bf 2.34} 13.15} & \scriptsize{{\bf 0.79  0.80}} & \scriptsize{{\bf 0.00  0.00}} & \scriptsize{{\bf 1.84  3.30}} & \scriptsize{{\bf 0.00  0.00}} & \scriptsize{{\bf 1.45  5.21}} \\
\hline
\scriptsize{2.23} & \scriptsize{PC1} $\eta$ \scriptsize{=0.025} & \scriptsize{PC2} $\eta$ \scriptsize{=0.025} & \scriptsize{99.85 99.94} & \scriptsize{99.89 99.94} & \scriptsize{99.94 99.94} & \scriptsize{99.77 99.94} & \scriptsize{99.91 99.94} & \scriptsize{99.81 99.94} \\
\scriptsize{2.24} & \scriptsize{PC1} $\eta$ \scriptsize{=0.05} & \scriptsize{PC2} $\eta$ \scriptsize{=0.05} & \scriptsize{99.82 99.94} & \scriptsize{99.94 99.94} & \scriptsize{99.94 99.94} & \scriptsize{99.87 99.94} & \scriptsize{99.88 99.94} & \scriptsize{99.94 99.94} \\
\scriptsize{2,25} & \scriptsize{PC1} $\eta$ \scriptsize{=0.3} & \scriptsize{PC2} $\eta$ \scriptsize{=0.3} & \scriptsize{99.69 99.94} & \scriptsize{99.89 99.94} & \scriptsize{99.94 99.94} & \scriptsize{98.39 99.94} & \scriptsize{99.94 99.94} & \scriptsize{99.87 99.94} \\
\scriptsize{2.26} & \scriptsize{PC1} $\eta$ \scriptsize{=0.4} & \scriptsize{PC2} $\eta$ \scriptsize{=0.4} & \scriptsize{55.01 99.94} & \scriptsize{62.69 99.91} & \scriptsize{13.21 16.31} & \scriptsize{65.91 99.94} & \scriptsize{30.13 37.76} & \scriptsize{83.58 99.94} \\
\hline
\hline
\multicolumn{7}{|l|}{\bf{Tests using {\sc starlab} and {\sc nbody6}}} \\
\scriptsize{3.1} & \scriptsize{standard} & \scriptsize{PC3/GPU} & \scriptsize{87.41 78.24} & \scriptsize{62.78 87.11} & \scriptsize{88.98 82.84} & \scriptsize{65.80 44.58} & \scriptsize{37.98 46.33} & \scriptsize{42.62 56.20} \\
\scriptsize{3.2} & \scriptsize{PC1} & \scriptsize{PC3} & \scriptsize{94.11 74.08} & \scriptsize{81.33 90.53} & \scriptsize{{\bf 1.05  1.29}} & \scriptsize{41.68 29.64} & \scriptsize{69.88 87.79} & \scriptsize{53.41 18.48} \\
\scriptsize{3.3} & \scriptsize{PC3/GPU} & \scriptsize{PC3} & \scriptsize{88.41 90.83} & \scriptsize{89.97 39.29} & \scriptsize{12.73 14.04} & \scriptsize{82.22 75.36} & \scriptsize{97.00 98.75} & \scriptsize{94.30 46.76} \\
\hline
\end{tabular}
\label{tab:res_boot_struct}
\end{center}
\end{table*}

\begin{table*}

\caption{Bootstrap results for energy-related cluster properties. Given are 
the fractions (in \%) of the test cluster distributions (which represent 
the case that they are drawn from the same parent distribution) more deviating than the main
simulations, i.e. the smaller this number the less alike the distributions are.
The ``standard'' setup uses PC1/GPU and an accuracy  setting of $\eta$=0.1.
Quantities are: 
E$_{\rm pot}$ = cluster potential energy, 
E$_{\rm kin}$ = cluster kinetic energy,
Q$_{vir}$ = cluster virial ratio,
E$_{\rm tot}$ = cluster total energy, 
$\delta$E$_{\rm tot}$ = change in cluster total energy during 1 $N$-body time unit. 
For each parameter the probabilities from the $\Delta_{12}$ and $\Gamma_{12}$
distributions (see Sect. \ref{sec:methods.timeseries}) are given. Note:  $^e$
denotes data sets with ``extreme'' uncertainties, i.e. ``extreme run-to-run''
scatter (applies only to the $\eta$=0.4 data sets). $^s$ denotes data sets with
strong ``spikes''/``wiggles'' in the median parameter evolution, compared to
other data sets in the same category.} 

\begin{center}
\begin{tabular}{l l l c c c c c}
\hline

\#ID & setup1 & setup2          & E$_{\rm pot}$ & E$_{\rm kin}$ & Q$_{vir}$ & E$_{\rm tot}$ & $\delta$E$_{\rm tot}$ \\
       &        &                & $\Delta_{12}$  $\Gamma_{12}$ & $\Delta_{12}$  $\Gamma_{12}$ & $\Delta_{12}$  $\Gamma_{12}$ & $\Delta_{12}$  $\Gamma_{12}$ & $\Delta_{12}$  $\Gamma_{12}$\\
\hline
\multicolumn{7}{|l|}{\bf{Tests using different PCs and accelerator hardware}} \\
1.1 & PC1/GPU   & PC1  	   & 36.33 74.09 & 78.07 94.07 & 98.78 85.97 & 73.78 79.77 & 10.97 67.01 \\
1.2 & PC1/GPU   & PC2/GRAPE	   & 57.97 99.21 & 38.76 43.44 & 41.61 39.07 & 67.69 53.08 & 12.28 39.33 \\
1.3 & PC2/GRAPE  & PC2  	   & {\bf 9.13} 24.42 & 25.44 28.05 & 42.35 43.88 & 31.91 34.91 & 41.07 95.35 \\
1.4 & PC1	 & PC2  	   & 78.35 99.94 & 99.91 99.94 & 99.94 99.94 & 83.79 99.94 & 36.01 99.86 \\
\hline
\hline
\multicolumn{7}{|l|}{\bf{Tests using different accuracy settings $\eta$}} \\
2.1 & standard & PC1/GPU $\eta$=0.05	 & {\bf 7.70  0.80} & 35.54 39.81 & 48.15 49.59 & {\bf 0.79  0.71} & {\bf 0.80  0.80} \\
2.2 & standard & PC1/GPU $\eta$=0.2	 & 77.95 83.90 & 69.85 79.04 & 88.87 80.26 & 81.35 86.65 & {\bf 1.07  1.87} \\
2.3 & standard & PC1/GPU $\eta$=0.3	 & 19.04  {\bf 7.42} & {\bf  0.36  0.40} & {\bf  0.79  0.85} &  {\bf 1.16  1.26} &  {\bf  0.80  0.80} \\
2.4 & standard & PC1/GPU $\eta$=0.4	 & {\bf 0.83  0.90} & 58.73 78.17 $^e$ & 61.43 95.12 $^e$ & 86.74 99.94 $^e$ & {\bf 0.68  0.80} \\
\hline
2.5 & PC1/GPU $\eta$=0.05  & PC1    $\eta$=0.05 & 15.92 {\bf  0.80}   & 88.13 61.33   & 87.25 94.55   & {\bf  0.44  0.36 }  & {\bf  0.80  0.80} \\
2.6 & PC1/GPU $\eta$=0.3   & PC1    $\eta$=0.3  & 33.84 24.99  & 50.99 26.35  & 27.13 16.48  &  60.81 58.50 & 13.87 {\bf  1.73} \\
2.7 & PC1/GPU $\eta$=0.4   & PC1    $\eta$=0.4  & {\bf  0.99  1.06}  & 84.21 99.94  $^e$ & 85.30 99.94  $^e$ &  95.44 99.94 $^e$ & 16.38  {\bf 2.15} \\
\hline
2.8 & PC1 $\eta$=0.1  & PC1 $\eta$=0.025 & 54.62 73.73 & 32.20 37.03  & 45.17 50.23  &  35.63 42.21 & {\bf  6.65} 78.87 \\
2.9 & PC1 $\eta$=0.1  & PC1 $\eta$=0.05  & 19.24 45.50  & 41.95 43.63  & 37.25 39.30  & 40.89 36.31  &  {\bf 9.72} 96.04 \\
2.10 & PC1 $\eta$=0.1  & PC1 $\eta$=0.3   & 56.43 33.79 & {\bf  0.26  0.32}	 & {\bf 0.63  0.67}  &  {\bf 1.28  1.39}  & {\bf 0.80  0.80} \\
2.11 & PC1 $\eta$=0.1  & PC1 $\eta$=0.4   & {\bf 7.82  1.39}  & 86.25 99.94 $^e$  & 87.80 99.94 $^e$ & 95.09 99.94 $^e$  & {\bf 0.78  0.80} \\
\hline
2.12 & PC2/GRAPE $\eta$=0.1 & PC2/GRAPE $\eta$=0.05  & 54.20 40.15  & {\bf  6.29  6.41}  & 13.68 11.89  & 19.97 12.78 & {\bf  4.32} 21.91  \\
2.13 & PC2/GRAPE $\eta$=0.1 & PC2/GRAPE $\eta$=0.2   &  41.02 77.53 &  13.73 14.83 & 10.71 10.76  & 18.60 18.18  &  {\bf 0.80  0.88} \\
2.14 & PC2/GRAPE $\eta$=0.1 & PC2/GRAPE $\eta$=0.25  &  29.41 25.22 & {\bf  0.36  0.38}  & {\bf  0.30  0.30}  & {\bf  1.35  1.47} &  {\bf 0.00  0.00}  \\
2.15 & PC2/GRAPE $\eta$=0.1 & PC2/GRAPE $\eta$=0.3   &  99.94 99.94 & {\bf  0.06  0.06}  & {\bf  0.58  0.62}  & {\bf  0.01  0.01} &  {\bf 0.00  0.00}  \\
\hline
2.16 & PC2 $\eta$=0.1 & PC2 $\eta$=0.025  & 42.86 74.61  & 29.10 34.59  & 40.68 46.22  & 24.94 30.01 & {\bf  0.78} 99.94  \\
2.17 & PC2 $\eta$=0.1 & PC2 $\eta$=0.05   & 12.26 22.60  & 41.97 43.63  & 37.21 39.24  & 47.95 40.23 & {\bf  4.93} 67.99  \\
2.18 & PC2 $\eta$=0.1 & PC2 $\eta$=0.2	  & 55.86 61.63 &  38.84 42.36 & 49.19 56.02 &  48.23 48.72 &  {\bf 0.80  0.81} \\
2.20 & PC2 $\eta$=0.1 & PC2 $\eta$=0.25   & 10.10 25.20 & 12.12 12.31 & 24.05 24.53  &  41.47 41.30 &  {\bf 0.00  0.00} \\
2.21 & PC2 $\eta$=0.1 & PC2 $\eta$=0.3	  & 40.47 33.49 & {\bf  0.26  0.32}  &  {\bf  0.63  0.67} &  {\bf  1.50  1.61} & {\bf  0.78  0.80}  \\
2.22 & PC2 $\eta$=0.1 & PC2 $\eta$=0.4    & {\bf 3.81  0.96} & {\bf 0.00  0.00} & {\bf 0.00  0.00} & {\bf 0.49  0.75} & {\bf 0.00  0.00} \\
\hline
2.23 & PC1 $\eta$=0.025 & PC2 $\eta$=0.025 & 99.84 99.94 & 99.91 99.94 & 99.93 99.94  & 99.88 99.94  & {\bf 0.73} 99.94  \\
2.24 & PC1 $\eta$=0.05 & PC2 $\eta$=0.05 & 99.72 99.94  & 99.90 99.94  & 99.79 99.94  & 99.78 99.94  & {\bf 5.98} 89.68  \\
2,25 & PC1 $\eta$=0.3  & PC2 $\eta$=0.3  & 99.75 99.94  & 99.91 99.94  & 99.92 99.94  &  99.94 99.94 &  74.42 99.94 \\
2.26 & PC1 $\eta$=0.4  & PC2 $\eta$=0.4  & 90.92 99.87  &  90.53 99.94 & 91.68 99.94  & 96.84 99.94  & 0.88  1.10 \\
\hline
\hline
\multicolumn{7}{|l|}{\bf{Tests using {\sc starlab} and {\sc nbody6}}} \\
3.1 & standard & PC3/GPU & 57.81 93.14 & 67.28 75.89 & 89.47 93.85 & 65.10 70.36 & {\bf 0.80  0.82} \\
3.2 & PC1      & PC3	 & 33.04 62.02 & 15.09 16.65 & 25.11 29.54 & 16.68 18.97 & {\bf 0.80  0.81} \\
3.3 & PC3/GPU  & PC3	 & 53.66 54.56 & 11.52 12.50 & 30.14 35.33 & 19.37 21.52 & {\bf 7.22  2.87} \\
\hline
\end{tabular}
\label{tab:res_boot_energy}
\end{center}
\end{table*}

\begin{table*}


\caption{Kuiper test results for binary properties. Given is the probability (in
\%) that for the given setups the binary properties are drawn from the same
distribution. The ``standard'' setup uses PC1/GPU and an accuracy setting of 
$\eta$=0.1. Quantities are: axis = semi-major axis, ecc = eccentricity, E/kT =
binding energy in E/kT, m2/m1 = mass ratio secondary mass / primary mass. \#X =
number of binaries in simulation X.} 

\begin{center}
\begin{tabular}{l l l c  c c  c c c c c c}
\hline
\#ID & setup1 & setup2 & \#1 & \#2 &          & axis & ecc & E/kT & m2/m1\\ 
\hline
\multicolumn{7}{|l|}{\bf{Tests using different PCs and accelerator hardware}} \\
1.1 & PC1/GPU       & PC1	  & 201 & 202 & 	& 83.57 & 83.99 & 60.39 & {\bf 4.99}\\
1.2 & PC1/GPU       & PC2/GRAPE  & 201 & 206 & 	& 80.18 & 32.28 & 69.14 & {\bf 1.63}\\
1.3 & PC2/GRAPE      & PC2	  & 206 & 202 & 	& 99.97 & 13.67  & 49.19 & 76.06\\
1.4 & PC1	     & PC2	  & 202 & 202 & 	& 100.00 & 100.00 & 100.00 & 100.00\\
\hline
\hline
\multicolumn{7}{|l|}{\bf{Tests using different accuracy settings $\eta$}} \\
2.1 & standard & PC1/GPU $\eta$=0.05   & 201 & 190 &	     & 88.96 & 38.18 & 42.01 & 97.89\\
2.2 & standard & PC1/GPU $\eta$=0.2    & 201 & 213 &	     & 97.26 & 72.84 & 90.72 & 44.90\\
2.3 & standard & PC1/GPU $\eta$=0.3    & 201 & 213 &	     & {\bf  9.20} & 51.32 &  {\bf 0.81} & 28.47 \\
2.4 & standard & PC1/GPU $\eta$=0.4    & 201 & 253 &	     &  {\bf 0.00} & 65.53 & {\bf 0.00} &  51.58\\
\hline
2.5 & PC1/GPU $\eta$=0.05  & PC1    $\eta$=0.05 &  190  & 202  &  & 29.28  &  77.83 & 53.79  & 33.28  \\
2.6 & PC1/GPU $\eta$=0.3   & PC1    $\eta$=0.3  &  213  & 220  &  & 58.73  &  96.36 & 56.65  & 32.64  \\
2.7 & PC1/GPU $\eta$=0.4   & PC1    $\eta$=0.4  &  253  & 260  &  & 67.64  &  85.16 & {\bf  0.12}  & 79.14  \\
\hline
2.8 & PC1    $\eta$=0.1  & PC1    $\eta$=0.025 & 202  & 204  &   & 94.15  & {\bf  6.24}  & 82.66  & 28.19  \\
2.9 & PC1    $\eta$=0.1  & PC1    $\eta$=0.05  & 202  & 202  &   & 20.43  & 37.05  & 58.81  & {\bf  2.33}  \\
2.10 & PC1    $\eta$=0.1  & PC1    $\eta$=0.3   & 202  & 220  &   & 20.24  & 18.05  & {\bf  1.10}  & {\bf  3.38} \\
2.11 & PC1    $\eta$=0.1  & PC1    $\eta$=0.4   & 202  & 260  &   & {\bf  0.00}  & 38.59  & {\bf  0.00}  & 21.45  \\
\hline
2.12 & PC2/GRAPE $\eta$=0.1 & PC2/GRAPE $\eta$=0.05    & 206  & 191  &	& 35.85  & 92.07   & 47.82 & 27.43  \\
2.13 & PC2/GRAPE $\eta$=0.1 & PC2/GRAPE $\eta$=0.2     & 206  & 204  &	& 43.81  & 64.46  & 42.24  &  17.05  \\
2.14 & PC2/GRAPE $\eta$=0.1 & PC2/GRAPE $\eta$=0.25    & 206  & 198  &	& 28.62  & 75.40  & 26.71  &  21.53  \\
2.15 & PC2/GRAPE $\eta$=0.1 & PC2/GRAPE $\eta$=0.3     & 206  & 234  &	& 48.59  & {\bf  4.71}   &  {\bf 5.57} & {\bf  4.23}  \\
\hline
2.16 & PC2    $\eta$=0.1 & PC2 $\eta$=0.025  & 202  & 204  &   &  93.99  & {\bf 8.03} & 82.66 & 28.19  \\
2.17 & PC2    $\eta$=0.1 & PC2 $\eta$=0.05   & 202  & 201  &   &  20.43  & 37.05 &  58.81 &  {\bf 2.33}  \\
2.18 & PC2    $\eta$=0.1 & PC2 $\eta$=0.2  & 202  & 204  &   &  64.77  &  {\bf 3.43} & 41.88  & 33.95  \\
2.20 & PC2    $\eta$=0.1 & PC2 $\eta$=0.25 & 202  & 202  &   &  66.44  & 30.82 & 73.82  &  {\bf 0.27}  \\
2.21 & PC2    $\eta$=0.1 & PC2 $\eta$=0.3  & 202  & 220  &   &  20.24  & 18.05 & {\bf 1.10} &  {\bf 3.38} \\
2.22 & PC2    $\eta$=0.1 & PC2    $\eta$=0.4 & 202  & 248  &   &  {\bf 0.03}   & 54.16 & {\bf 0.00} & 19.66  \\
\hline
2.23 & PC1    $\eta$=0.025 & PC2 $\eta$=0.025   & 204  & 204  &   &  100.00 & 100.00  & 100.00  &  100.00 \\
2.24 & PC1    $\eta$=0.05  & PC2    $\eta$=0.05 & 202  & 202  &   &  100.00 & 100.00  & 100.00  &  100.00 \\
2,25 & PC1    $\eta$=0.3   & PC2    $\eta$=0.3  & 220  & 220  &   &  100.00 & 100.00  & 100.00  &  100.00 \\
2.26 & PC1    $\eta$=0.4   & PC2    $\eta$=0.4  & 260  & 248  &   & 93.34   & 99.88   &  {\bf 0.22}   & 99.97  \\
\hline
\hline
\multicolumn{7}{|l|}{\bf{Tests using {\sc starlab} and {\sc nbody6}}} \\
3.1 & standard & PC3/GPU & 201 & 189 &        & 48.81 & 12.63 & 54.46 & 76.14\\
3.2 & PC1      & PC3	 & 202 & 209 &        & 46.91 & 34.27 & 74.56 & {\bf 7.80}\\
3.3 & PC3/GPU  & PC3	 & 189 & 209 &        & 22.69 & 43.63 & 80.84 & 49.64\\
\hline
\end{tabular}
\label{tab:kstest_bin}
\end{center}
\end{table*}

\begin{table*}

\caption{Kuiper test results for energy-related properties of unbound stars at 
1000 $N$-body time units (i.e. well after core collapse). Results with a "B"  
superscript denote results for unbound binary systems. Given is the probability
(in \%) that for the given setups the properties are drawn from the same
distribution. The ``standard'' setup uses PC1/GPU acceleration
and an accuracy setting of  $\eta$=0.1. Quantities are: E$_{\rm kin}$ = reduced
kinetic energy, E$_{\rm pot}$ = reduced potential energy, E$_{\rm tot}$ =
reduced total energy. \#X = number of systems in simulation X. } 

\begin{center}
\begin{tabular}{l l l  l l  c c c l lc c c}
\hline
\#ID & setup1 & setup2              & \#1 & \#2 &   E$_{\rm kin}$ & E$_{\rm pot}$ & E$_{\rm tot}$ & \#1$^B$ & \#2$^B$ & E$^B_{\rm kin}$ & E$^B_{\rm pot}$ & E$^B_{\rm tot}$\\
\hline
\multicolumn{7}{|l|}{\bf{Tests using different PCs and accelerator hardware}} \\
\scriptsize{1.1} & \scriptsize{PC1/GPU} & \scriptsize{PC1  } & \scriptsize{7554} & \scriptsize{7441} & \scriptsize{{\bf4.71}} & \scriptsize{31.42} & \scriptsize{13.47} & \scriptsize{144} & \scriptsize{150} & \scriptsize{96.68} & \scriptsize{94.73} & \scriptsize{96.46} \\
\scriptsize{1.2} & \scriptsize{PC1/GPU} & \scriptsize{PC2/GRAPE} & \scriptsize{7554} & \scriptsize{7518} & \scriptsize{11.47} & \scriptsize{11.08} & \scriptsize{34.37} & \scriptsize{144} & \scriptsize{142} & \scriptsize{86.53} & \scriptsize{87.59} & \scriptsize{86.53} \\
\scriptsize{1.3} & \scriptsize{PC2/GRAPE} & \scriptsize{PC2  } & \scriptsize{7518} & \scriptsize{7441} & \scriptsize{49.20} & \scriptsize{73.55} & \scriptsize{59.18} & \scriptsize{142} & \scriptsize{150} & \scriptsize{84.42} & \scriptsize{62.13} & \scriptsize{91.79} \\
\scriptsize{1.4} & \scriptsize{PC1} & \scriptsize{PC2  } & \scriptsize{7441} & \scriptsize{7441} & \scriptsize{100.00} & \scriptsize{100.00} & \scriptsize{100.00} & \scriptsize{150} & \scriptsize{150} & \scriptsize{100.00} & \scriptsize{100.00} & \scriptsize{100.00} \\
\hline
\hline
\multicolumn{7}{|l|}{\bf{Tests using different accuracy settings $\eta$}} \\
\scriptsize{2.1} & \scriptsize{standard} & \scriptsize{PC1/GPU} $\eta$ \scriptsize{=0.05} & \scriptsize{7554} & \scriptsize{7592} & \scriptsize{30.86} & \scriptsize{16.44} & \scriptsize{62.53} & \scriptsize{144} & \scriptsize{144} & \scriptsize{86.44} & \scriptsize{92.10} & \scriptsize{86.44} \\
\scriptsize{2.2} & \scriptsize{standard} & \scriptsize{PC1/GPU} $\eta$ \scriptsize{=0.2} & \scriptsize{7554} & \scriptsize{7640} & \scriptsize{21.75} & \scriptsize{16.36} & \scriptsize{34.70} & \scriptsize{144} & \scriptsize{154} & \scriptsize{86.56} & \scriptsize{36.04} & \scriptsize{74.98} \\
\scriptsize{2.3} & \scriptsize{standard} & \scriptsize{PC1/GPU} $\eta$ \scriptsize{=0.3} & \scriptsize{7554} & \scriptsize{8199} & \scriptsize{{\bf 0.00}} & \scriptsize{{\bf 0.00}} & \scriptsize{{\bf 0.00}} & \scriptsize{144} & \scriptsize{160} & \scriptsize{88.99} & \scriptsize{67.80} & \scriptsize{93.17} \\
\scriptsize{2.4} & \scriptsize{standard} & \scriptsize{PC1/GPU} $\eta$ \scriptsize{=0.4} & \scriptsize{7554} & \scriptsize{10052} & \scriptsize{{\bf 0.00}} & \scriptsize{{\bf 0.00}} & \scriptsize{{\bf 0.00}} & \scriptsize{144} & \scriptsize{209} & \scriptsize{15.77} & \scriptsize{35.88} & \scriptsize{30.41} \\
\hline
\scriptsize{2.5} & \scriptsize{PC1/GPU} $\eta$ \scriptsize{=0.05} & \scriptsize{PC1} $\eta$ \scriptsize{=0.05} & \scriptsize{7592} & \scriptsize{7557} & \scriptsize{90.44} & \scriptsize{88.07} & \scriptsize{87.85 } & \scriptsize{144} & \scriptsize{136} & \scriptsize{13.86} & \scriptsize{95.14} & \scriptsize{17.95} \\
\scriptsize{2.6} & \scriptsize{PC1/GPU} $\eta$ \scriptsize{=0.3} & \scriptsize{PC1} $\eta$ \scriptsize{=0.3} & \scriptsize{8199} & \scriptsize{7364} & \scriptsize{{\bf 0.00}} & \scriptsize{{\bf 0.00}} & \scriptsize{{\bf 0.00}} & \scriptsize{160} & \scriptsize{169} & \scriptsize{64.47} & \scriptsize{71.74} & \scriptsize{47.64} \\
\scriptsize{2.7} & \scriptsize{PC1/GPU} $\eta$ \scriptsize{=0.4} & \scriptsize{PC1} $\eta$ \scriptsize{=0.4} & \scriptsize{10052} & \scriptsize{9567} & \scriptsize{{\bf 0.00}} & \scriptsize{{\bf 0.11}} & \scriptsize{{\bf 0.00}} & \scriptsize{209} & \scriptsize{197} & \scriptsize{82.27} & \scriptsize{32.84} & \scriptsize{74.65} \\
\hline
\scriptsize{2.8} & \scriptsize{PC1} $\eta$ \scriptsize{=0.1} & \scriptsize{PC1} $\eta$ \scriptsize{=0.025} & \scriptsize{7441} & \scriptsize{7659} & \scriptsize{36.43} & \scriptsize{55.25} & \scriptsize{16.48} & \scriptsize{150} & \scriptsize{148} & \scriptsize{60.58} & \scriptsize{32.59} & \scriptsize{69.61} \\
\scriptsize{2.9} & \scriptsize{PC1} $\eta$ \scriptsize{=0.1} & \scriptsize{PC1} $\eta$ \scriptsize{=0.05} & \scriptsize{7441} & \scriptsize{7557} & \scriptsize{16.38} & \scriptsize{64.68} & \scriptsize{22.13} & \scriptsize{150} & \scriptsize{136} & \scriptsize{65.41} & \scriptsize{91.57} & \scriptsize{65.41} \\
\scriptsize{2.10} & \scriptsize{PC1} $\eta$ \scriptsize{=0.1} & \scriptsize{PC1} $\eta$ \scriptsize{=0.3} & \scriptsize{7441} & \scriptsize{7364} & \scriptsize{{\bf 3.71}} & \scriptsize{{\bf 7.26}} & \scriptsize{12.84} & \scriptsize{150} & \scriptsize{169} & \scriptsize{54.09} & \scriptsize{84.83} & \scriptsize{72.21} \\
\scriptsize{2.11} & \scriptsize{PC1} $\eta$ \scriptsize{=0.1} & \scriptsize{PC1} $\eta$ \scriptsize{=0.4} & \scriptsize{7441} & \scriptsize{9567} & \scriptsize{{\bf 0.00}} & \scriptsize{{\bf 0.00}} & \scriptsize{{\bf 0.00}} & \scriptsize{150} & \scriptsize{197} & \scriptsize{15.87} & \scriptsize{{\bf 6.38}} & \scriptsize{15.87} \\
\hline
\scriptsize{2.12} & \scriptsize{PC2/GRAPE} $\eta$ \scriptsize{=0.1} & \scriptsize{PC2/GRAPE} $\eta$ \scriptsize{=0.05} & \scriptsize{7518} & \scriptsize{7405} & \scriptsize{98.19} & \scriptsize{50.21} & \scriptsize{97.57} & \scriptsize{142} & \scriptsize{134} & \scriptsize{99.92} & \scriptsize{67.48} & \scriptsize{99.96} \\
\scriptsize{2.13} & \scriptsize{PC2/GRAPE} $\eta$ \scriptsize{=0.1} & \scriptsize{PC2/GRAPE} $\eta$ \scriptsize{=0.2} & \scriptsize{7518} & \scriptsize{7493} & \scriptsize{71.42} & \scriptsize{80.02} & \scriptsize{74.28} & \scriptsize{142} & \scriptsize{151} & \scriptsize{49.01} & \scriptsize{49.97} & \scriptsize{49.01} \\
\scriptsize{2.14} & \scriptsize{PC2/GRAPE} $\eta$ \scriptsize{=0.1} & \scriptsize{PC2/GRAPE} $\eta$ \scriptsize{=0.25} & \scriptsize{7518} & \scriptsize{7366} & \scriptsize{41.45} & \scriptsize{88.32} & \scriptsize{48.86} & \scriptsize{142} & \scriptsize{149} & \scriptsize{56.28} & \scriptsize{46.16} & \scriptsize{62.32} \\
\scriptsize{2.15} & \scriptsize{PC2/GRAPE} $\eta$ \scriptsize{=0.1} & \scriptsize{PC2/GRAPE} $\eta$ \scriptsize{=0.3} & \scriptsize{7518} & \scriptsize{7254} & \scriptsize{{\bf 8.61}} & \scriptsize{22.26} & \scriptsize{{\bf 5.48}} & \scriptsize{142} & \scriptsize{161} & \scriptsize{30.29} & \scriptsize{23.97} & \scriptsize{30.29} \\
\hline
\scriptsize{2.16} & \scriptsize{PC2} $\eta$ \scriptsize{=0.1} & \scriptsize{PC2} $\eta$ \scriptsize{=0.025} & \scriptsize{7441} & \scriptsize{7659} & \scriptsize{36.43} & \scriptsize{55.25} & \scriptsize{17.08} & \scriptsize{150} & \scriptsize{148} & \scriptsize{60.58} & \scriptsize{32.59} & \scriptsize{69.61} \\
\scriptsize{2.17} & \scriptsize{PC2} $\eta$ \scriptsize{=0.1} & \scriptsize{PC2} $\eta$ \scriptsize{=0.05} & \scriptsize{7441} & \scriptsize{7557} & \scriptsize{16.38} & \scriptsize{64.68} & \scriptsize{22.13} & \scriptsize{150} & \scriptsize{136} & \scriptsize{65.41} & \scriptsize{91.57} & \scriptsize{65.41} \\
\scriptsize{2.18} & \scriptsize{PC2} $\eta$ \scriptsize{=0.1} & \scriptsize{PC2} $\eta$ \scriptsize{=0.2} & \scriptsize{7441} & \scriptsize{7414} & \scriptsize{{\bf  3.52}} & \scriptsize{38.18} & \scriptsize{{\bf 5.98}} & \scriptsize{150} & \scriptsize{145} & \scriptsize{80.71} & \scriptsize{53.26} & \scriptsize{73.67} \\
\scriptsize{2.20} & \scriptsize{PC2} $\eta$ \scriptsize{=0.1} & \scriptsize{PC2} $\eta$ \scriptsize{=0.25} & \scriptsize{7441} & \scriptsize{7497} & \scriptsize{{\bf  6.38}} & \scriptsize{84.60} & \scriptsize{24.08} & \scriptsize{150} & \scriptsize{158} & \scriptsize{99.50} & \scriptsize{92.74} & \scriptsize{99.91} \\
\scriptsize{2.21} & \scriptsize{PC2} $\eta$ \scriptsize{=0.1} & \scriptsize{PC2} $\eta$ \scriptsize{=0.3} & \scriptsize{7441} & \scriptsize{7364} & \scriptsize{{\bf 3.71}} & \scriptsize{{\bf 7.26}} & \scriptsize{12.84} & \scriptsize{150} & \scriptsize{169} & \scriptsize{54.09} & \scriptsize{84.83} & \scriptsize{72.21} \\
\scriptsize{2.22} & \scriptsize{PC2} $\eta$ \scriptsize{=0.1} & \scriptsize{PC2} $\eta$ \scriptsize{=0.4} & \scriptsize{7441} & \scriptsize{7398} & \scriptsize{{\bf 0.00}} & \scriptsize{{\bf 0.00}} & \scriptsize{{\bf 0.00}} & \scriptsize{150} & \scriptsize{190} & \scriptsize{{\bf 2.74}} & \scriptsize{{\bf 6.40}} & \scriptsize{{\bf 2.74}} \\
\hline
\scriptsize{2.23} & \scriptsize{PC1} $\eta$ \scriptsize{=0.025} & \scriptsize{PC2} $\eta$ \scriptsize{=0.025} & \scriptsize{7659} & \scriptsize{7659} & \scriptsize{100.00} & \scriptsize{100.00} & \scriptsize{100.00} & \scriptsize{148} & \scriptsize{148} & \scriptsize{100.00} & \scriptsize{100.00} & \scriptsize{100.00} \\
\scriptsize{2.24} & \scriptsize{PC1} $\eta$ \scriptsize{=0.05} & \scriptsize{PC2} $\eta$ \scriptsize{=0.05} & \scriptsize{7557} & \scriptsize{7557} & \scriptsize{100.00} & \scriptsize{100.00} & \scriptsize{100.00} & \scriptsize{136} & \scriptsize{136} & \scriptsize{100.00} & \scriptsize{100.00} & \scriptsize{100.00} \\
\scriptsize{2.25} & \scriptsize{PC1} $\eta$ \scriptsize{=0.3} & \scriptsize{PC2} $\eta$ \scriptsize{=0.3} & \scriptsize{7364} & \scriptsize{7364} & \scriptsize{100.00} & \scriptsize{100.00} & \scriptsize{100.00} & \scriptsize{169} & \scriptsize{169} & \scriptsize{100.00} & \scriptsize{100.00} & \scriptsize{100.00} \\
\scriptsize{2.26} & \scriptsize{PC1} $\eta$ \scriptsize{=0.4} & \scriptsize{PC2} $\eta$ \scriptsize{=0.4} & \scriptsize{9567} & \scriptsize{7398} & \scriptsize{{\bf 0.00}} & \scriptsize{{\bf 0.00}} & \scriptsize{{\bf 0.00}} & \scriptsize{197} & \scriptsize{190} & \scriptsize{99.91} & \scriptsize{91.22} & \scriptsize{99.99} \\
\hline
\hline
\multicolumn{7}{|l|}{\bf{Tests using {\sc starlab} and {\sc nbody6}}} \\
\scriptsize{3.1} & \scriptsize{standard} & \scriptsize{PC3/GPU} & \scriptsize{7554} & \scriptsize{7501} & \scriptsize{{\bf 9.59}} & \scriptsize{48.69} & \scriptsize{21.44} & \scriptsize{144} & \scriptsize{135} & \scriptsize{99.84} & \scriptsize{99.22} & \scriptsize{99.94} \\
\scriptsize{3.2} & \scriptsize{PC1} & \scriptsize{PC3} & \scriptsize{7441} & \scriptsize{7507} & \scriptsize{81.64} & \scriptsize{17.46} & \scriptsize{63.79} & \scriptsize{150} & \scriptsize{146} & \scriptsize{92.75} & \scriptsize{71.24} & \scriptsize{87.68} \\
\scriptsize{3.3} & \scriptsize{PC3/GPU} & \scriptsize{PC3} & \scriptsize{7501} & \scriptsize{7507} & \scriptsize{72.60} & \scriptsize{37.09} & \scriptsize{82.08} & \scriptsize{135} & \scriptsize{146} & \scriptsize{79.35} & \scriptsize{29.11} & \scriptsize{85.84} \\
\hline
\end{tabular}
\label{tab:kstest_esc}
\end{center}
\end{table*}

\end{document}